\pdfoutput=1
\documentclass[journal=jctcce,manuscript=article]{achemso}
\setkeys{acs}{articletitle = true, doi=true}
\usepackage[version=3]{mhchem} % Formula subscripts using \ce{}
\usepackage[utf8]{inputenc}
\usepackage[T1]{fontenc}       % Use modern font encodings
\usepackage{color}
\usepackage{spreadtab}
\usepackage{algorithm,algorithmic}

\usepackage{fp} % floating point calculations
\usepackage{bm} % complies with mds tex
\usepackage{amssymb} % complies with mds tex

\author{Matteo De Santis}
\affiliation[Univ. Lille]
{Univ. Lille, CNRS, UMR 8523-PhLAM-Physique des Lasers Atomes et Mol\'{e}cules, F-59000 Lille, France}
\author{Diego Sorbelli}
\affiliation[Universit\`a degli Studi di Perugia]
{Dipartimento di Chimica, Biologia e Biotecnologie, Universit\`a degli Studi di Perugia,
Via Elce di Sotto 8, 06123
Perugia, Italy}
\alsoaffiliation{Istituto di Scienze e Tecnologie Chimiche (SCITEC), Consiglio Nazionale delle Ricerche
 c/o
Dipartimento di Chimica, Biologia e Biotecnologie, Universit\`a degli Studi di Perugia,
Via Elce di Sotto 8, 06123
Perugia, Italy}
\author{Val\'{e}rie Vallet}
\affiliation[Univ. Lille]
{Univ. Lille, CNRS, UMR 8523-PhLAM-Physique des Lasers Atomes et Mol\'{e}cules, F-59000 Lille, France}
\author{Andr\'{e} Severo Pereira Gomes}
\affiliation[Univ. Lille]
{Univ. Lille, CNRS, UMR 8523-PhLAM-Physique des Lasers Atomes et Mol\'{e}cules, F-59000 Lille, France}
\author{Loriano Storchi}
\email{loriano@storchi.org}
\affiliation[Universit\`a degli Studi `G. D'Annunzio']
{Dipartimento di Farmacia, Universit\`a degli Studi `G. D'Annunzio',
Via dei Vestini 31, 66100
Chieti, Italy}
\alsoaffiliation{Istituto di Scienze e Tecnologie Chimiche (SCITEC), Consiglio Nazionale delle Ricerche
 c/o
Dipartimento di Chimica, Biologia e Biotecnologie, Universit\`a degli Studi di Perugia,
Via Elce di Sotto 8, 06123
Perugia, Italy}
\author{Leonardo Belpassi}
\email{leonardo.belpassi@cnr.it}
\affiliation[Consiglio Nazionale delle Ricerche]
{Istituto di Scienze e Tecnologie Chimiche (SCITEC), Consiglio Nazionale delle Ricerche
 c/o
Dipartimento di Chimica, Biologia e Biotecnologie, Universit\`a degli Studi di Perugia,
Via Elce di Sotto 8, 06123
Perugia, Italy}

\title{Frozen-Density Embedding for including environmental effects in the Dirac-Kohn-Sham theory:
an implementation based on density fitting and prototyping techniques}

\begin{document}
\maketitle

\begin{abstract}
The Frozen Density Embedding (FDE) scheme represents an embedding method in which
environmental effects onto a given subsystem are included by representing the other subsystems
making up the surroundings quantum mechanically, by means of their electron densities. In the
present paper, we extend the full 4-component relativistic Dirac-Kohn-Sham (DKS) method, as
implemented in the BERTHA code, to include environmental and confinement effects with the FDE
scheme (DKS-in-DFT FDE).  This implementation has been enormously facilitated by BERTHA's python
API (PyBERTHA), which provides a flexible framework of development by using all Python advantages in terms of code re-usability, portability while facilitating
the interoperability with other FDE implementations available through the PyADF framework.  The accuracy and numerical stability
of this new implementation, also using different auxiliary fitting basis sets,
has been demonstrated on the simple NH$_3$-H$_2$O system in comparison with 
a reference non relativistic implementation.
The computational performance has been evaluated on a series of gold 
clusters (Au$_n$, with $n=2,4,8$) embedded into an increasing number 
of water molecules (5, 10, 20, 40 and 80 water molecules).  
We found that the procedure scales
approximately linearly both with the size of the frozen surrounding environment (in line with the underpinnings of the FDE approach) and
with the size of the active system (in line with the use of density fitting). 
Finally, we applied the code to a series of Heavy (Rn)
and Super-Heavy elements (Cn, Fl, Og) embedded in a C$_{60}$ cage to
explore the confinement effect induced by C$_{60}$ 
on their electronic structure. We compare the results from our simulations with more approximate models employed in the atomic physics literature, in which confinement is represented by a radial potential slightly affected by the nature of the central atom. Our results indicate that the specific interactions described by FDE are able to improve upon the cruder approximations currently employed, and thus provide a basis from which to generate more realistic radial potentials for confined atoms.  
\end{abstract}

\section{Introduction}

Molecular systems, clusters, and materials containing heavy atoms have
drawn considerable recent attention because of their rich 
chemistry and physics\cite{Schwerdtfeger2020,pyykko2012relativistic,Pershina2013,Giuliani2019}. In
order to model computationally systems containing heavy elements, the
methods of relativistic quantum mechanics must be necessarily adopted to capture
scalar and spin-orbit interactions that are
neglected in the conventional non-relativistic formulation of quantum
chemistry.  
Furthermore, most of the chemistry occurs in solution and 
the environment plays a key role in the determination of the properties 
and reactivity of substances in condensed phases\cite{Orozco2000,Maher2012,Kumpulainen2016,Gerber2020,Dupuy2021}.
%Thus, the methods of (relativistic) quantum chemistry needs to be extended 
Thus, the complexity of chemical phenomena in solution has made it necessary to develop 
a variety of models and computational techniques to be combined with (relativistic) quantum chemistry methods.
Among different approaches to include environmental effects we mention
the quantum mechanics/molecular mechanics (QM/MM) approach~\cite{Warshel1976}, which includes the molecular environment explicitly and at a reduced cost using 
classical mechanical description, or in polarizable continuous medium (PCM)\cite{Tomasi2005} 
(i.e., where the solvent degree of freedom are replaced by an effective classical dielectric).
Despite being widely and successfully applied these methods may have drawbacks.
For instance methods based on PCM cannot describe specific interactions with the environments (e.g. hydrogen, halogen bonds), 
while the QM/MM approach, which is based on classical force fields, may be limited by the availability of accurate 
parameterizations which may reduce its predictive
power, in particular when heavy elements are involved. 
An alternative is to use quantum embedding theories (for an overview see Ref.\citenum{Huang2008,SeveroPereiraGomes2012,Sun2016,Jones2020}
and references therein), 
in which a QM description for a subsystem of interest is combined with a QM description of the environment (QM/QM).
A notable example of QM/QM methods is the frozen-density 
embedding (FDE) scheme  introduced by Weso\l{}owski and 
Warshel,\cite{Wesolowski1993,Wesolowski2015} based on the approach originally
proposed by Senatore and Subbaswamy~\cite{Senatore1986}, and later Cortona~\cite{Cortona1992},
for solid-state calculations. The method has been further
generalized~\cite{Iannuzzi2006,jacob2008flexible} and directed to the simultaneous
optimization of the subsystems electronic densities.

FDE is a DFT-in-DFT embedding method that allows to partition a larger
Kohn-Sham system into a set of smaller, and coupled, Kohn-Sham subsystems.
The coupling term is defined by a local embedding potential 
depending only on the electron densities of both the sole active subsystem and the environment (i.e. no orbital information is shared among subsystems).
This feature gives to the FDE scheme an enormous flexibility, as indeed 
virtually arbitrary methods can be combined to treat different subsystems.
For example, wavefunction theory (WFT) based methods can be used for the active system
while one can take advantage of the efficiency of DFT to describe 
a large environment (WFT-in-DFT)~\cite{Kluner2002,Govind1998,Huang2006,Huang2008,SeveroPereiraGomes2012,Dresselhaus2015,Wesolowski2015}.
Also one can employ very different computational protocols for different subsystem 
including i) using Hamiltonian dealing with different  relativistic approximations
(from the full 4-component methods to the non-relativistic ones)~\cite{Gomes2008a,Gtz2014,Olejniczak2017,Halbert2020}; 
ii) different basis sets size and type (Gaussian and Slater type functions, relativistic 
four-component spinors) 
and even iii) different quantum chemical packages\cite{Gomes2008a,Jacob2011a,DeSantis2020b}. 
We mention that the FDE scheme  has been extended both 
to the linear-response TDDFT~\cite{Casida2004,Neugebauer2007,Neugebauer2009}, 
including to account for charge-transfer excitations~\cite{Tolle2019,Tolle2019b} 
and to real time TDDFT (rt-TDDFT).\cite{Krishtal2015,DeSantis2020b}

FDE-based calculations are shown to be accurate in the
case of  weakly interacting systems including hydrogen bond systems\cite{Fux2010,Gotz2009}, 
whereas,
their use for subsystems interacting with a larger covalent character
is problematic (see Ref.\citenum{Fux2010} and references therein) due 
to the use of approximate kinetic energy functional (KEDF) in the non-additive
contribution to the embedding potential.
The research for more accurate KEDFs is a key aspect for the
applicability of the FDE scheme as a general scheme~\cite{Goodpaster2010,Constantin2018,Jiang2018,Constantin2019}, including the partitioning
of the system also breaking covalent bonds.~\cite{Mi2020}
We mention here that alternative QM/QM approaches,
avoiding the use of KEDFs and allowing also for fragmentation in 
subsystems through covalent bonds,
have been recently proposed (see for instance Ref.\cite{Manby2012,Cohen2007,Elliott2010,Nafziger2011,Hgely2016,Sun2016,Lee2019,Jones2020,Mosquera2020}).
%For instance, the Manby-Miller scheme involves the projection techniques, and allows the mapping of
%different domains within the total system into two different-quality levels (basis set size or
%different exchange-correlation functional)
%employed in each domain, however an extension to tract different domains beyond 
%single-determinant  approximation (Hartree-Fock or DFT),
%or different Hamiltonian for including relativistic effects at different levels, 
%is still unexplored and may be not straightforward.

Thanks to its flexibility the FDE scheme has been implemented in different 
flavors into computational packages  such as: embedded 
Quantum Espresso\cite{Genova2017}, ADF\cite{ADF2017authors,jacob2008flexible,Johannes:2005}, Turbomole~\cite{Laricchia2010,Laricchia2013}, Dalton~\cite{Gomes2008a,Hfener2013}, Koala~\cite{Hfener2020}, Molpro~\cite{Manby2012},  Serenity\cite{Unsleber2018a}, and Q-Chem~\cite{Epifanovsky2020x},
(the first two based on plane waves and Slater type functions respectively, the others on Gaussian type functions).
FDE has also been implemented to treat 
the subsystems at full-relativistic four-component level based on the
Dirac equation within the DIRAC code\cite{Saue2020}, and can be used with DFT and different wavefunction methods both 
for molecular properties and energies involving the ground or excited electronic states~\cite{Gomes2008a,Hfener2012,Gomes2013,Olejniczak2017,Bouchafra2018,Halbert2020}.

Despite its conceptual simplicity, its actual implementations may lead 
to relatively complicated workflows. A simpler approach is therefore to integrate such legacy codes as computational engines to handle the different 
FDE steps, which are then glued together and their execution automatized using suitable frameworks such as for instance 
that implemented in PyADF \cite{Jacob2011a,pyadf-github-v096-python3},
that can be easily extensible due to its object-oriented 
implementation in the Python programming language~\cite{van1995python}.
Prototyping  techniques also based on Python are very useful to build 
reference implementations,
as for instance the \textsc{Psi4-rt-PyEmbed} code\cite{DeSantis2020b,pyberthagit}, where
the Python interface of Psi4Numpy and PyADF~\cite{Jacob2011a,pyadf-github-v096}
(including its PyEmbed module\cite{pyembed,SchmittMonreal2020} and XCFun
library\cite{Ekstrom2010,xcfun:2019} to evaluate non-additive exchange-correlation
and kinetic energy contributions) has been used by some of us
 to build real time non-relativistic TDDFT-in-DFT FDE~\cite{DeSantis2020b} and projection-based embedding\cite{DeSantis2022} implementations.

In this work we extend the Dirac-Kohn-Sham (DKS) method implemented in the  BERTHA code (with its
new Python API, PyBERTHA)\cite{parcopaper,Belpassi2020} 
to the FDE scheme 
to include environmental/confinement effects in the DKS calculations (DKS-in-DFT FDE).
The implementation takes advantages of the DKS formulation implemented in BERTHA, 
including the density fitting algorithms 
at the core of the computation (i.e. in the evaluation of the embedding potential and of 
its matrix representation is relativistic G-spinor functions),
and the FDE implementation already available  in the PyEmbed module of the 
PyADF framework.

The outline of the paper is as follows. In Section 2 we present the basic theory
of FDE and a brief description of the DKS method
as implemented in BERTHA.
In section 3, we then describe in detail our implementation.  
In section 4 we present some numerical results,
including the computational burden and scalability of this new implementation
with respect to the size of the active system as well as of the embedding one.
We will also present an application  to a series of Heavy (Rn)
and Super-Heavy elements (Cn, Fl, Og) confined into a C$_{60}$ cage.
Finally, concluding remarks are given in section 5.

\section{Theory}

In this section, we briefly review the basic formalism of the  FDE scheme 
and its extension to use the DKS theory  for the active system 
(DKS-in-DFT FDE).
We will remark also some details of the DKS implementation in BERTHA,
mainly focusing on those aspects (including density fitting techniques),
which are relevant for an efficient implementation of the FDE scheme.
Finally, we will illustrate the basic characteristics
of the our recent BERTHA Python API, PyBERTHA (and the related \textbf{pyberthamod} module
available under GPLv3 license at Ref.\citenum{pyberthagit}, for additional and technical
details see Refs.\citenum{DeSantis2020a,parcopaper,Belpassi2020}) which is a key tool here to devise a
simple work-flow for the DKS-in-DFT FDE scheme.

\subsection{Subsystem DFT and Frozen Density Embedding formulation}

In the subsystem formulation of DFT the entire system is partitioned into N subsystems,
and the total density $\rho_\text{tot}(\bm{r})$ is represented as the sum of electron densities of the
various subsystems [i.e., $\rho_a(\bm{r})$ ($a = I,..,N$)].
In the following we consider the total density as partitioned in only two
contributions as
\begin{equation}
        \rho_\text{tot}(\bm{r}) =\rho_\text{I}(\bm{r}) + \rho_\text{II}(\bm{r}).
\end{equation}
The total energy of the system can then be written as
\begin{equation}
        E_\text{tot}[\rho_\text{I},\rho_\text{II}] = E_\text{I}[\rho_\text{I}] + E_\text{II}[\rho_\text{II}] + E_\text{int}[\rho_\text{I},\rho_\text{II}]
\label{eq:etot}
\end{equation}
with the energy of each subsystem ($E_i[\rho_i]$, with $i=\text{I},\text{II}$) given according to the usual
definition in DFT as
\begin{equation}
\begin{aligned}
E_i[\rho_i]   &= \int\rho_i(\bm{r})v_\text{nuc}^{i}(\bm{r}) {\rm d}^3r
                          + \frac{1}{2}\iint\frac{\rho_i(\bm{r})\rho_{i}(\bm{r}')}{|\bm{r}-\bm{r}'|}{\rm d}^3r {\rm d}^3r' + \\
              &+ E_\text{xc}[\rho_i] + T_s[\rho_i] + E_\text{nuc}^{i}.
\end{aligned}
\end{equation}
In the above expression, $v_\text{nuc}^{i}(\bm{r})$ is the nuclear potential due to the set of atoms which defines the
subsystem, and $E_\text{nuc}^{i}$ is the related nuclear repulsion energy.
$T_s[\rho_i]$ is the kinetic energy of the auxiliary non-interacting system,
which is, within the Kohn-Sham (KS) approach, commonly evaluated using the KS orbitals.
The interaction energy is given by the expression:
\begin{equation}
\begin{aligned}
  \label{eq:eint}
        E_\text{int}[\rho_\text{I},\rho_\text{II}] &= \int\rho_\text{I}(\bm{r})v_\text{nuc}^\text{II}(\bm{r}){\rm d}^3r
                    +\int\rho_\text{II}(\bm{r})v_\text{nuc}^\text{I}(\bm{r}) {\rm d}^3r + E^\text{I,II}_\text{nuc} \\
        &+\iint\frac{\rho_\text{I}(\bm{r})\rho_\text{II}(\bm{r}')}{|\bm{r}-\bm{r}'|} {\rm d}^3r {\rm d}^3r'
                   +E^\text{nadd}_\text{xc}[\rho_\text{I},\rho_\text{II}] + T^\text{nadd}_s[\rho_\text{I},\rho_\text{II}]
\end{aligned}
\end{equation}
with $v_\text{nuc}^\text{I}$ and $v_\text{nuc}^\text{II}$ being the nuclear potentials due to the set of atoms associated with the subsystem $\text{I}$
and $\text{II}$,
respectively. The repulsion energy for nuclei belonging to different subsystems is described by the $E^\text{I,II}_\text{nuc}$ term.
The non-additive contributions ($E^\text{nadd}_\text{xc}[\rho_\text{I},\rho_\text{II}]$ and $T^\text{nadd}_s[\rho_\text{I},\rho_\text{II}])$ arise because both exchange-correlation and kinetic energy, in contrast to the Coulomb interaction,
are not linear functionals of the density.

The electron density of a given fragment ($\rho_\text{I}$ or $\rho_\text{II}$ in this case)
can be determined by minimizing the total energy
functional (Eq.\ref{eq:etot}) with respect to the density of the fragment while
keeping the density of the other subsystem frozen. This procedure is the essence of the
FDE scheme and leads to a set of Kohn-Sham-like equations (one for  each subsystem)
\begin{equation}\label{eq:act_opt}
        \Big[ \mathcal{T} + v^\text{KS}_\text{eff}[\rho_\text{I}](\bm{r})
                + v_\text{emb}^\text{I}[\rho_\text{I},\rho_\text{II}](\bm{r})\Big]\phi_k^\text{I}(\bm{r}) = \varepsilon_k^\text{I}\phi_k^\text{I}(\bm{r}).
\end{equation}
which are coupled by the embedding potential term $v^\text{I}_\text{emb}(\bm{r})$,
which carries all dependence on the other fragment's density. Here $\mathcal{T}$ denotes the kinetic energy operator, which in a non-relativistic framework has the form $-\nabla^2/2$, whereas for a relativistic framework is $c \bm{\alpha}
\cdot {\bf p}$ (see discussion below). We also note that in the relativistic framework, the FDE expressions above correspond to the case in which an external vector potential is absent. Further details for their generalization can be found in Ref.~\citenum{Olejniczak2017}.

In this equation, $v^\text{KS}_\text{eff}[\rho_\text{I}](\bm{r})$ is the KS potential calculated on basis
of the density of subsystem $\text{I}$ only, whereas $v_\text{emb}^\text{I}[\rho_\text{I},\rho_\text{II}](\bm{r})$ is the embedding potential that takes into account the effect of the other subsystem
(which we consider here as the complete environment).
In the framework of FDE theory, $v^\text{I}_\text{emb}(\bm{r})$ is explicitly given by
\begin{equation}\label{eq:vemb}
\begin{aligned}
        v^\text{I}_\text{emb}[\rho_\text{I},\rho_\text{II}](\bm{r})
        = \frac{\delta E_\text{int}[\rho_\text{I},\rho_\text{II}]}{\delta\rho_\text{I}(\bm{r})}=&
            \ v_\text{nuc}^\text{II}(\bm{r})+\int\frac{\rho_\text{II}(\bm{r}')}{|\bm{r}-\bm{r}'|} {\rm d}^3r'
          + \frac{\delta E_\text{xc}^\text{nadd}[\rho_\text{I}, \rho_\text{II}]}{\delta\rho_\text{I}(\bm{r})}
          + \frac{\delta T^\text{nadd}_s[\rho_\text{I}, \rho_\text{II}]}{\delta\rho_\text{I}(\bm{r})},
\end{aligned}
\end{equation}
where the non-additive exchange-correlation and kinetic energy contributions
are defined as the difference between the associated exchange-correlation and the kinetic
potentials defined using $\rho_\text{tot}(\bm{r})$ (i.e., $\rho_\text{I}(\bm{r})+\rho_\text{II}(\bm{r})$) and $\rho_\text{I}(\bm{r})$. For both potentials
one needs to account for the fact that only the density is known for the total system, so that
potentials that require input in the form of KS orbitals are prohibited.
For the exchange-correlation potential, one may make use of accurate density functional
approximations and its quality is therefore similar to that of ordinary KS. The potential for the non-additive kinetic term
($\frac{\delta T^\text{nadd}_s[\rho]}{\delta\rho_\text{I}(\bm{r})}$,
in Eq.\ref{eq:vemb}) is more problematic as less accurate orbital-free kinetic energy density functionals (KEDFs)
are available for this purpose.
Examples of popular functional approximations applied in this context are the Thomas-Fermi (TF)
kinetic energy functional\cite{Thomas1927} or
the GGA functional  PW91k\cite{Lembarki1994}.
As already mentioned in the introduction, the research for more accurate KEDFs is a key aspect for the
applicability of the FDE scheme as a general scheme, including the partitioning
of the system also breaking covalent bonds.\cite{Mi2020}

In general, the set of coupled equations that arise in the FDE scheme for the
 subsystems  have to be solved iteratively and a freeze-and-thaw scheme,
 where one relaxes the electron density of one subsystem at a time  keeping
frozen the others, until electron densities of all subsystems reach 
a required convergence.  
In this work we limit to one subsystem (active)
while keeping the density of the environment frozen to their ground state density.
In this case, the implementation of FDE reduces to the evaluation 
of $v^\text{I}_\text{emb}(\bm{r})$ potential  (which is an one-electron operator) that
needs to be added to the Hamiltonian of the active system.
The matrix representation of the embedding potential
may be evaluated using numerical integration grids,
similar to those used for the exchange-correlation term in the KS method.
This contribution is then added  to the KS matrix and the eigenvalue problem
is solved with the usual self-consistent field (SCF) procedure.
We note here that two approaches can be taken: the first is the use of a pre-calculated embedding potential~\cite{Gomes2008a} (for instance, from a prior subsystem DFT calculation) that is used as a one-body operator (referred to in the literature as a ``static'' embedding potential) added to the one-body Fock matrix at the start of the (4-component) calculations. The second approach involves the regeneration of the embedding potential using the (4-component) actual electron density of the active system. In this case, the matrix representation of the embedding potential is updated during SCF procedure, because of its dependence on the active subsystem density (see Eq.\ref{eq:vemb}) that itself  changes during the SCF iterations. As discussed below, in this work we shall mostly make use of the latter approach.

\subsection{Dirac-Kohn-Sham scheme in BERTHA and its extension to FDE based 
on density fitting}

For the detailed theoretical basis of the Dirac-Kohn-Sham methodology
we refer the reader to previous works\cite{Liu1997, Varga2000, Yanai2001a,
Saue2002, Komorovsky2008, Zhang2020, Repisky2020} and references therein.
Here, we summarize only the main aspects of the implementation of an all-electron DKS method based on the use
of G-spinor basis sets and  the density-fitting techniques 
as is implemented in BERTHA\cite{Belpassi2011, Belpassi2020}. 
In atomic units, and including only the longitudinal electrostatic potential,
the DKS equation reads \begin{equation}\label{eq:dks} \{c \bm{\alpha}
\cdot {\bf p}+ {\beta} c^2+v^{(l)}({\bf r})\} {\bm \Psi}_{i}({\bf
r})=\varepsilon_i{\bm \Psi}_{i}({\bf r}), \end{equation} where $c$
is the speed of light in vacuum, $\bf p$ is the electron
momentum, while:
\begin{equation}
{\bm\alpha} = \left(
\begin{array}{cc}
0 & {\bm\sigma} \\
{\bm\sigma} & 0
\end{array}
\right)\ \mbox{and}\
{\bm\beta} =
\left(
\begin{array}{cc}
{\bm I} & 0 \\
0 & -{\bm I}
\end{array}
\right)
\end{equation}
where ${\bm\sigma}=(\sigma_x,\sigma_y,\sigma_z)$, $\sigma_q$ is a
$2\times 2$ Pauli spin matrix and ${\bm I}$
is the $2\times 2$ identity matrix.
The longitudinal interaction term is represented by a diagonal operator borrowed from
non-relativistic theory and made up of: a nuclear potential term $v_{\mathrm{N}}(\bm{r})$,
a Coulomb interaction term $v^{(l)}_{\mathrm{H}}[\rho(\bm{r})]$,
and the exchange-correlation term $v^{(l)}_{\mathrm{XC}}[\rho(\bm{r})]$.
We mention that the Breit interaction contributes to the transverse part of the Hartree
interaction and is not considered here, as we restrict ourselves to using non-hybrid, non-relativistic functionals of the electron density.

In BERTHA, the spinor solution (${\bm \Psi}_{i}({\bf r})$ in 
Eq.~\ref{eq:dks}) is expressed as a linear combination of the G-spinor
basis functions \cite{grant2007relativistic}, $M_\mu^T(\bm{r})$ ($T=L,S$ with
$L$ and $S$ refer to the so-called ``large'' and ``small'' component,
respectively).  The G-spinors do not suffer from the variational problems
of kinetic balance (see Ref.~\citenum{Dyall1990} and references
therein) and,  regarding the evaluation of multicentre integrals,
retain the advantages that have made Gaussian-type functions the
most widely-used expansion set in non-relativistic quantum chemistry.
The matrix representation of the DKS operator in the G-spinor basis is
given by
\begin{align}\label{eq:H_DKS}
\bm{H}_{DKS} &= \begin{bmatrix}
                      \bm{V}^{(LL)} +mc^2 \bm{S}^{(LL)} & c\bm{\Pi}^{(LS)}  \\
                                   c\bm{\Pi}^{(SL)}      &  \bm{V}^{(SS)} -mc^2\bm{S}^{(SS)}
                     \end{bmatrix},
\end{align}
where
\begin{equation}\label{eq:V_DKS}
\bm{V}^{(TT)} = \bm{v}^{(TT)} + \bm{J}^{(TT)} + \bm{K}^{(TT)}.
\end{equation}
The eigenvalue equation in the algebraic representation is given by
\begin{equation}
\bm{H}_{DKS}\begin{bmatrix} \bm{c}^{(L)} \\
                             \bm{c}^{(S)}
                              \end{bmatrix} = E
                              \begin{bmatrix} \bm{S}^{(LL)} & 0 \\
                                0 & \bm{S}^{(SS)}
                                \end{bmatrix}
                                \begin{bmatrix}
                                   \bm{c}^{(L)}\\
                                   \bm{c}^{(S)} \end{bmatrix}
\end{equation}
where $\bm{c}^{(T)}$ are the spinor expansion vectors. The $\bm{H}_{DKS}$ matrix is defined
in terms of the $\bm{v}^{(TT)}, \bm{J}^{(TT)}, \bm{K}^{(TT)}, \bm{S}^{(TT)}$, and $\bm{\Pi}^{(TT')}$ matrices,
being respectively the basis representation of the nuclear, Coulomb, and exchange-correlation potentials, the overlap matrix, and
the matrix of the kinetic operator, respectively.
The nuclear charges have been modeled by a finite
Gaussian distribution~\cite{Visscher1997}.

The  resulting matrix elements are defined by
\begin{align}
  & \bm{v}_{\mu \nu}^{(TT)} = \int \bm{v}_{N}(\bm{r})\rho_{\mu \nu}^{(TT)}(\bm{r})d\bm{r}   \\
  & J_{\mu \nu}^{(TT)} = \int v_{H}^{(\mathrm{l})}[\rho(\bm{r})]\rho_{\mu \nu}^{(TT)}(\bm{r})d\bm{r}\label{eq:ref1}   \\
  & K_{\mu \nu}^{(TT)} = \int v_{\mathrm{xc}}^{(\mathrm{l})}[\rho(\bm{r})]\rho_{\mu \nu}^{(TT)}(\bm{r})d\bm{r}\label{eq:ref2}  \\
  & S_{\mu \nu}^{(TT)} = \int \rho_{\mu \nu}^{(TT)}(\bm{r})d\bm{r} \\
  & \Pi_{\mu \nu}^{TT'} = \int M_\mu^{(T) \dagger}(\bm{r})(\bm{\sigma}\cdot \bm{p}) M_\nu  ^{(T')}(\bm{r})d\bm{r} .
\end{align}
The terms $\rho_{\mu\nu}^{(TT)}({\bf r})$ are the G-spinor overlap densities ($M^{(T) \dagger}_{\mu}({\bf r}) M^{(T)}_{\nu}({\bf r}) $),
which  can be exactly expressed as
linear combination of standard Hermite Gaussian-type
functions (HGTFs).\cite{Grant2000,grant2007relativistic,Belpassi2011}
The $\bm{H}_{DKS}$ matrix depends on $\rho(\bm{r})$
in $v_{\mathrm{xc}}^{(\mathrm{l})}[\rho(\bm{r})]$
and $v_{\mathrm{H}}^{(\mathrm{l})}[\rho(\bm{r})]$, through the canonical spinors obtained by
its diagonalization. Thus, the solutions
$\bm{c}^{(T)}$ are solved self-consistently.

In the G-spinor representation, we define the density matrix ($\bm{D}_{TT'}$)
as the product column by row of the $c_\mu^{(T)}$ coefficients
($D^{TT'}_{\mu \nu} = \sum_i c^{(T)*}_{\mu i} c^{(T')}_{\nu i}$,
with $T$ and $T'$ equal to both $L$ and $S$), where the sum runs over
the occupied positive-energy states.  The total electron density is
obtained according to \begin{equation}\label{eq:tot_el_den} \rho(\bm{r}) =
\sum_{T} \sum_{\mu, \nu} D_{\mu \nu}^{(TT)} \rho_{\mu \nu}^{(TT)}(\bm{r}).
\end{equation}
The computation of the Coulomb and exchange-correlation contributions
to the DKS matrix, that is Eq. \ref{eq:ref1} and \ref{eq:ref2}, respectively,
is the most demanding computational step in a DKS calculation involving
a G-spinor basis set.
The current version of BERTHA takes advantage of both density fitting 
\cite{Belpassi2006,Belpassi2008,Belpassi2008a,Belpassi2011}
and of advanced parallelization techniques\cite{Storchi2010, Storchi2013, Rampino2014, Belpassi2020,parcopaper} for the evaluation of these two 
contributions.
The relativistic density (which is a real scalar function) is thereby expanded in a set 
of N$_{\mathrm{aux}}$ auxiliary atom-centered functions.

\begin{equation}\label{eq:fitted_dens}
\tilde{\rho} = \sum_{t=1}^{N_{aux}} d_tf_t(\bm{r}).
\end{equation}

In the Coulomb metric the expansion coefficients $d_t$ are defined as the solution
of the linear system, given by

\begin{equation}\label{eq:A_mat}
\bm{A}\bm{d} = \bm{v},
\end{equation}

where $\bm{A}$ is a real and symmetric matrix, representing the Coulomb
interaction in the auxiliary basis, $A_{st} = \langle f_s\,||\,f_t
\rangle$ while the elements ($v_s$) of the vector $\bm{v}$ are the
projection of the electrostatic potential on the fitting functions,
$\langle f_s\,||\,\rho \rangle$. 
In our implementation the latter integrals are efficiently evaluated using the
relativistic generalization of the scalar Hermite density matrix proposed
by Alml\"of \cite{RezaAhmadi1995, Challacombe1996}.

For the exchange-correlation $\bm{K}$ matrix we adopt a
similar strategy by solving for $\bm{z}$ in the linear system
\begin{equation}\label{eq:lin_sys} \bm{A}\bm{z} = \bm{w},
\end{equation} where the vector $\bm{w}$ is the projection
of the exchange-correlation potential ($\tilde{v}^{(\mathrm{l})}_{\mathrm{xc}}[\tilde{\rho}(\bm{r})]$)
on the fitting functions 
\begin{equation} w_s = \langle
\tilde{v}^{(\mathrm{l})}_{\mathrm{xc}} | f_s \rangle = \int
\tilde{v}^{(\mathrm{l})}_{\mathrm{xc}}[\tilde{\rho}(\bm{r})]f_s(\bm{r})d\bm{r}.
\end{equation} 
The elements of the vector ${\bf w}$, which involve
integrals of the exchange-correlation potential, are computed
numerically by the integration scheme already implemented in the
code.\cite{Quiney2002}  Once the vectors $\bm{d}$ and $\bm{z}$ 
have
been worked out, the Coulomb and the exchange-correlation contributions
to the DKS matrix can be evaluated in terms of 3-center two-electron
integrals $I^{(TT)}_{s,\mu \nu} = \langle f_s\,||\,\rho_{\mu
\nu}^{(TT)}\rangle$: 
\begin{equation} 
\tilde{J}_{\mu \nu}^{(TT)} + \tilde{K}_{\mu \nu}^{(TT)} = \sum_{t=1}^{N_{\mathrm{aux}}} I_{t,\mu
\nu}^{(TT)}(d_t +z_t) 
\end{equation}

The extension of this scheme to include the contribution FDE potential,
$v^\text{I}_\text{emb}[\rho_\text{I},\rho_\text{II}](\bm{r})$,
is straightforward.  
For the evaluation of the embedding potential matrix representation
in G-spinors ($\tilde{V}^{emb(TT)}_{\mu \nu}$) we can strictly follow the procedure already 
employed above for the exchange-correlation term.
We solve for $\bm{c}$ in the linear system
\begin{equation}\label{eq:lin_sysemb}
\bm{A}\bm{c} = \bm{g},
\end{equation}
where the vector $\bm{g}$ is the projection of the embedding potential,
$v^\text{I}_\text{emb}[\rho_\text{I},\rho_\text{II}](\bm{r})$, on the fitting functions
\begin{equation}\label{eq:embproj}
g_s = \langle \tilde{v}_{\mathrm{emb}} | f_s \rangle = \int \tilde{v}_{\mathrm{emb}}[\tilde{\rho_I},\rho_{II}(\bm{r})]f_s(\bm{r})d\bm{r}.
\end{equation}
The elements of the vector ${\bf g}$ are computed numerically on a
suitable integration grid (further details of the implementation will be
given in the next section).  Once the vector $\bm{g}$ has been  computed
the embedding potential contribution can be evaluated in a single step 
together with the Coulomb and the exchange-correlation ones (see Eq.\ref{eq:embedcont}), and finally added to
the DKS matrix. 
\begin{equation}\label{eq:embedcont} \tilde{J}_{\mu
	\nu}^{(TT)} + \tilde{K}_{\mu \nu}^{(TT)} + \tilde{V}_{\mu \nu}^{emb(TT)} =
\sum_{t=1}^{N_{\mathrm{aux}}} I_{t,\mu \nu}^{(TT)}(d_t +z_t+ c_t)
\end{equation}
Note that, in the evaluation of
the embedding potential on a numerical grid,
$\tilde{v}_{\mathrm{emb}}(\bm{r})={v}_{\mathrm{emb}}[\tilde{\rho_I}(\bm{r}),\rho_{II}(\bm{r})]$,
we use the fitted density of the active system, $\tilde{\rho_I}$
(see Eq.\ref{eq:fitted_dens}). 
%This procedure is fully consistent with the
%variational fitting scheme already employed in BERTHA. 
An important
aspect for the computational efficiency arises from the use of
an auxiliary fitting basis set
of primitive HGTFs. Indeed, they are grouped together in sets sharing the same
exponents \cite{Belpassi2006}.  In particular, each set is defined so
that a specified auxiliary function of a given angular momentum is
associated with all the corresponding functions of smaller angular
momentum sharing the same exponent.  This allows us to use the polynomial Hermite
recurrence relations both in the analytical evaluation of the
two-electron integrals for the Coulomb term and in the numerical
representation of fitting basis functions used in the exchange-correlation
and embedding potential contributions. This further reduces the burden of the
expensive operation of evaluating large numbers of Gaussian exponents
at each grid point~\cite{Belpassi2008a}.

\section{Implementation and Computational Details}

%\subsection{The PyBERTHA API and software stack} 

Before addressing the FDE implementation, we present a brief outline the new Python \cite{van1995python} API that has been recently implemented and that contributed to
improve both the usability and interoperability of the BERTHA code\cite{parcopaper, DeSantis2020a,Belpassi2020}. 
All these  new features have been extensively employed in the work-flow design and 
implementation of the DKS-in-DFT FDE method (see next section).  
\begin{figure}[h!]
  \includegraphics[width=0.8\linewidth]{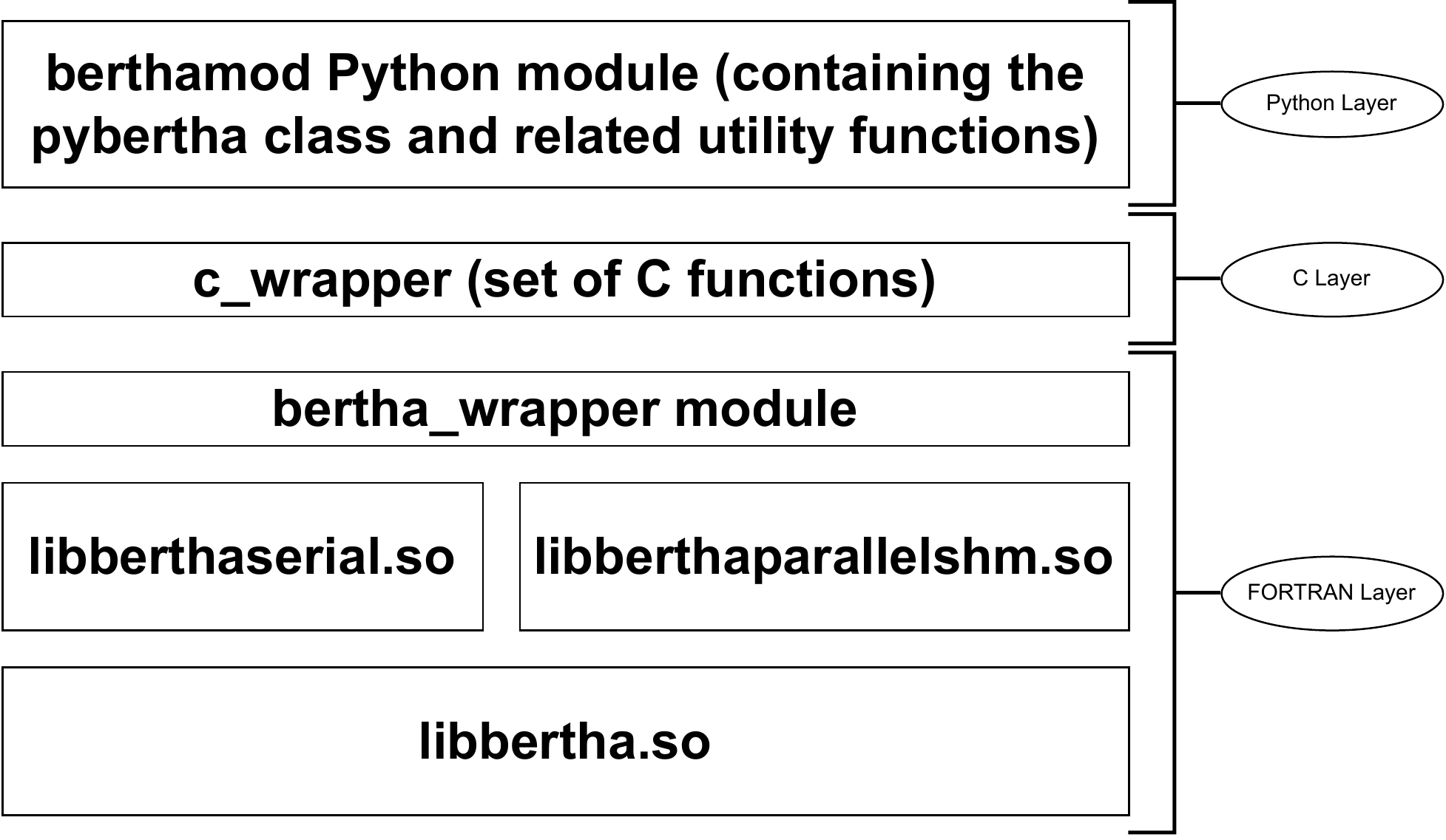}
  \caption{An overview of the BERTHA software's layers.}
  \label{fig:bsl}
\end{figure}
In Figure \ref{fig:bsl} we outline the fundamental structure of BERTHA. 
All the basic kernel functions written in FORTRAN are now collected in a single Shared Object (SO) 
(i.e, \textbf{libertha.so}). Alongside there are two other SO libraries:
\textbf{libberthaserial.so} capable of performing both the serial and parallel OpenMP based~\cite{Dagum1998} runs, and
\textbf{libberthaparalleshm.so} containing  all the functions needed to perform
MPI~\cite{clarke1994mpi} based parallel computations where also the memory burden is distributed among the processes.

We also implemented a FORTRAN module, named \textbf{bertha\_wrapper}, containing a class implementing all the methods 
needed to access to all the basic quantities, such as:
energy,  density, DKS and overlap matrices and other. 
The same FORTRAN
module (i.e., \textbf{bertha\_wrapper}) is used to perform all the basic operations such as:
\textbf{bertha\_init} to perform all the memory allocations, \textbf{bertha\_main} to run the main SCF iterations,
and \textbf{bertha\_finalize} to free all the allocated memory, and more.
Finally the main PyBERTHA \cite{pyberthagit} module has been developed using the \textbf{ctypes}
Python module. This module provides the C-compatible data types, and allows calling functions collected in
shared libraries. In order to simplify the direct interlanguage communication between Python and FORTRAN, we
implemented a simple C layer called \textbf{c\_wrapper}, also summarized in Figure~\ref{fig:bsl}.
This Python API to BERTHA has been  described in detail in Refs.\citenum{parcopaper,DeSantis2020a}
and in the present work  has been further extended with new methods which allow us
to extract  all those  quantities necessary for the DKS-in-DFT FDE implementation 
(e.g., the method \textbf{bertha\_get\_density\_on\_grid()} 
used to extract the values of the fitted electron density on a grid). 
All the new methods have been efficiently parallelized using OpenMP~\cite{Dagum1998}.
All details and computational efficiency will be given in the next sections.

\subsection{The PyBerthaEmbed DKS-in-DFT FDE implementation}

In this section we outline the computational strategy we adopted to
implement the DKS-in-DFT FDE scheme. 
The developed Python program \textbf{pyberthaemb.py} and the related module (\textbf{pyembmod}) are freely 
available under GPLv3 license at Ref.~\citenum{pyberthagitrefrel}.
A data set collection of computational results including numerical data, parameters and job input instructions used to obtain the results of 
Section \ref{sec:results}, is available and can be freely accessed at the Zenodo repository, see Ref. ~\citenum{matteo_de_santis_2022_6343894}.

\subsubsection{Implementation strategy}

%\begin{algorithm}[H]
\begin{algorithm}
\begin{algorithmic}[1]
\STATE \textbf{import berthamod}
\STATE \textbf{import pyembmod}
\STATE \textbf{...}
\STATE \textbf{bertha = berthamod.pybertha(pberthaopt.wrapperso)}
\STATE \textbf{\# set options for the DKS calculation}
\STATE \textbf{bertha.set\_fnameinput(fnameinput)}
\STATE \textbf{bertha.set\_fittfname(fittfname)}
\STATE \textbf{...}
\STATE \textbf{bertha.init()}
\STATE \textbf{ovapm, eigem, fockm, eigen = bertha.run()}
\STATE \textbf{...}
\STATE \textbf{activefname = pberthaopt.activefile}
\STATE \textbf{envirofname = pberthaopt.envirofile}
\STATE \textbf{embfactory = pyembmod.pyemb(activefname,envirofname,'adf') \#jobtype='adf' is default}
\STATE \textbf{embfactory.set\_options(param=pberthaopt.param,....) \# several paramenters to be specified in input- e.g AUG/ADZP for ADF}
\STATE \textbf{\# embfactory.set\_grid\_filename(pberthaopt.gridfname)  \# a general grid may be used}
\STATE \textbf{...}
\STATE \textbf{embfactory.initialize()}
\STATE \textbf{grid = embfactory.get\_grid()}
\STATE \textbf{...}
\STATE \textbf{rho = bertha.get\_density\_on\_grid(grid)}
\STATE \textbf{density=numpy.zeros((rho.shape[0],10))}
\STATE \textbf{density[:,0] = rho}
\STATE \textbf{...}
\STATE \textbf{pot = embfactory.get\_potential(density)}
\STATE \textbf{...}
\STATE \textbf{for out\_iter in range (maxiter): \# split-SCF scheme iterations, see text}
\STATE \textbf{    bertha.init()}
\STATE \textbf{    ...}
\STATE \textbf{    bertha.set\_embpot\_on\_grid(grid, pot)}
\STATE \textbf{    ovapm, eigem, fockm, eigen = bertha.run(eigem)}
\STATE \textbf{    rho = bertha.get\_density\_on\_grid(grid)}
\STATE \textbf{    density=numpy.zeros((rho.shape[0],10))}
\STATE \textbf{    density[:,0] = rho}
\STATE \textbf{    pot\_old=pot }
\STATE \textbf{    pot = embfactory.get\_potential(density)}
\STATE \textbf{    norm\_pot = numpy.sqrt(numpy.sum((pot-pot\_old)**2))}
\STATE \textbf{    norm\_D = numpy.linalg.norm(diffD,'fro')}
\STATE \textbf{    norm\_D = numpy.linalg.norm(diffD,'fro')}
\STATE \textbf{    if (norm\_D<(1.0e-6) and norm\_pot <(1.0e-4)):}
\STATE \textbf{        bertha.finalize()}
\STATE \textbf{        break}
\end{algorithmic}
  \caption{Illustrative Python code to compute active system DKS density (using the \textbf{pyberthaemb.py} code), 
	environment density and Coulomb potential (using the ADF code) and non-additive embedding potential via the \textbf{pyemb} module}
\label{maincode}
\end{algorithm}

%Thanks to our recent development of PyBERTHA (see above), 
%the implementation of the DKS-in-DFT FDE method (\textsc{PyBERTHAembed} software) resulted straightforward and relatively simple.
%We also took advantages of the software development already put in place by some of us
%in the \textsc{Psi4-rt-PyEmbed} code\cite{pybertha,pyberthagit}, where
%we used the Python interface of $Psi4Numpy$ and $PyADF$\cite{pyadf,pyadf-github-v096} 
%(including its $PyEmbed$ module\cite{pyembed,schmittmonreal_frozen-density_2020} and XCFun
%library\cite{ruud2010,xcfun:2019}) to build a real time TDDFT-in-DFT FDE implementation.
%The object-oriented programming (i.e. extensibility and inheritance)
%based of Python to build-in  classes  with an high degree of interoperability which 
%permitted one to handle different aspects and quantities
%involved in the work-flow also coming from different codes as a single unit and in a common framework based on Python.
%Here we adopt a very similar strategy. 
The newly developed code is composed of two main modules: 
the \textbf{pyembmod} one,
that allows to manage all the important quantities for the
FDE implementation, and the \textbf{pyberthamod} module \cite{pyberthagit}.

Specifically, the \textbf{pyemb} class inside the \textbf{pyembmod} module allows to well isolate all the 
FDE data and operations increasing the level of abstraction.
The module is used to manage all the required quantities for the generation of 
the embedding potential, that is ${v}_{\mathrm{emb}}[\tilde{\rho_I},\rho_{II}(\bm{r})]$. 
It has been engineered in a such manner that all details of the FDE low-lying implementation
will be completely transparent from the PyBERTHA side. This has the advantage 
that all future developments and/or integration of the FDE scheme
(g.e. using DKS theory also for the environment DKS-in-DKS FDE) will not
affect the \textsc{PyBERTHAembed}
code, i.e. it will remain completely unchanged.  In particular in this
first version the \textbf{pyembmod} module can handle the basic
procedures previously implemented in the \textsc{Psi4-rt-PyEmbed} software which
are based  on the use of PyADF\cite{Jacob2011a,pyadf-github-v096}, PyEmbed module\cite{pyembed,SchmittMonreal2020}, and the XCFun
library\cite{Ekstrom2010,xcfun:2019} to evaluate non-additive
exchange-correlation and kinetic energy contributions on a user-defined
integration grids.  This approach  gave us both the advantage of the
code re-usability and, even more importantly, a DFT-in-DFT FDE reference
implementation in which we can have the precise control over all those
details and parameters from which a FDE scheme depends on (i.e., algorithms,
numerical grid definition, quantum chemistry packages used to determine
electronic density and Coulomb potential of the environment, basis sets,
exchange-correlation functionals, etc...). This has clearly made the
debugging phase in the development of \textsc{PyBERTHAembed} software straightforward.

Algorithm \ref{maincode}  reports the most important part of the \textbf{pyberthaemb.py}
code, and it well illustrates how we can gain a relatively simple work-flow
to implement FDE using the DKS level of theory for the active system
using PyBERTHA and the new \textbf{pyemb} class for the environmental system.
The \textbf{pybertha} class is instantiated (line 4) with the
shared object \textbf{bertha\_wrapper.so} specified as an input. The SO contains the cited 
\textbf{c\_wrapper} and \textbf{bertha\_wrapper} code which are based on the
core FORTRAN libraries, namely: \textbf{libbertha.so} and \textbf{libberthaserial.so} (see Figure \ref{fig:bsl}
above).  After the initialization, the full DKS calculation is worked out 
(line 10) using the \textbf{bertha.run()} method. 
At line 14 the \textbf{pyemb} class is also instantiated specifying the files
(specified in xyz format) for the geometries of both the active and embedding
systems. The quantum chemistry software employed for the actual
calculation of the
environment system is specified at this stage (in the current example, and throughout this work, we 
used the ADF package\cite{ADF2017authors}). 

All the details for the computation of the embedding system are set at line 15.
This includes: the selection of the type of basis set functions, Hamiltonian,
exchange-correlation functional and also the
non-additive kinetic functional used to define the embedding potential.  
At this stage all the basis sets and exchange-correlation functionals available 
in the ADF library can be used. Similarly, the numerical integration grid 
used for the numerical representation of the embedding potential is set, 
both the type (global grid or active system) and the quality.
By default the numerical grids internally defined by the ADF program are used, however other options 
are available, including the possibility to use an user-defined grid 
(see for instance line 16, commented).

The \textbf{embfactory.initialize()} method performs  
a stand-alone single point calculation on the embedding system.
The method evaluates the nuclear and Coulomb potentials of the environment (see Eq.\ref{eq:vemb}) and its
ground state electron density ($\rho_{II}$). All these quantities are mapped on the numerical grid. 
At this time, the numerical grid defined within PyADF is made available (as a \textbf{numpy.array}) 
using the \textbf{get\_grid()} method (line 19)
and used as an input for the \textbf{get\_density\_on\_grid()} method of the
\textbf{pybertha} class.  This method allows to define the ground state density $\tilde \rho_{I}$ 
of the active system at DKS level of theory.
$\tilde \rho_{I}$ is also available as a \textbf{numpy.array} that, after a reshape (line 22), 
can be used as an input of the \textbf{get\_potential()} method of the \textbf{pyemb} class (line 25)
to obtain the final embedding potential.
%The embedding potential is thus calculated from its constituents
%(i.e. the environment electrostatic and nuclear potential and the non-additive contribution as
%detailed in Eq.~\ref{eq:vemb}) and evaluated at each grid point, $v^{emb}({r}_k)$.

After this initial setup, we proceed to the actual FDE calculation. In this example, the embedding potential will be generated using the active subsystem density (loop structure, lines 27 to 40), using the split-SCF scheme~\cite{Duak2009}. Thus, in each of the spin-SCF iterations, the new \textbf{set\_embpot\_on\_grid()} method of \textbf{pybertha} class makes 
both the numerical grid and the embedding potential available at the FORTRAN layer. Thus, the numerical integration 
of the $v^{emb}({r})$ on the fitting basis functions (Eq.\ref{eq:embproj}) and 
linear system solution (Eq.\ref{eq:lin_sysemb}) are efficiently evaluated in FORTRAN.
The DKS matrix employed in the intervening BERTHA
calculation (\textbf{bertha.run()}, line 31) 
is updated using the G-spinor representation of the embedding potential, and here the split-SCF scheme is interesting as it does not require the evaluation of the embedding potential at each SCF step taking place on the BERTHA side. 
The new fitted density (line 32) is used to compute a new embedding potential (line 36)
which is used in for the next iteration of the split-SCF procedure~\cite{Duak2009}. This scheme is iterated till 
a convergence criteria is satisfied (line 40). 

\begin{figure}[!h]
        \includegraphics[width=1\textwidth]{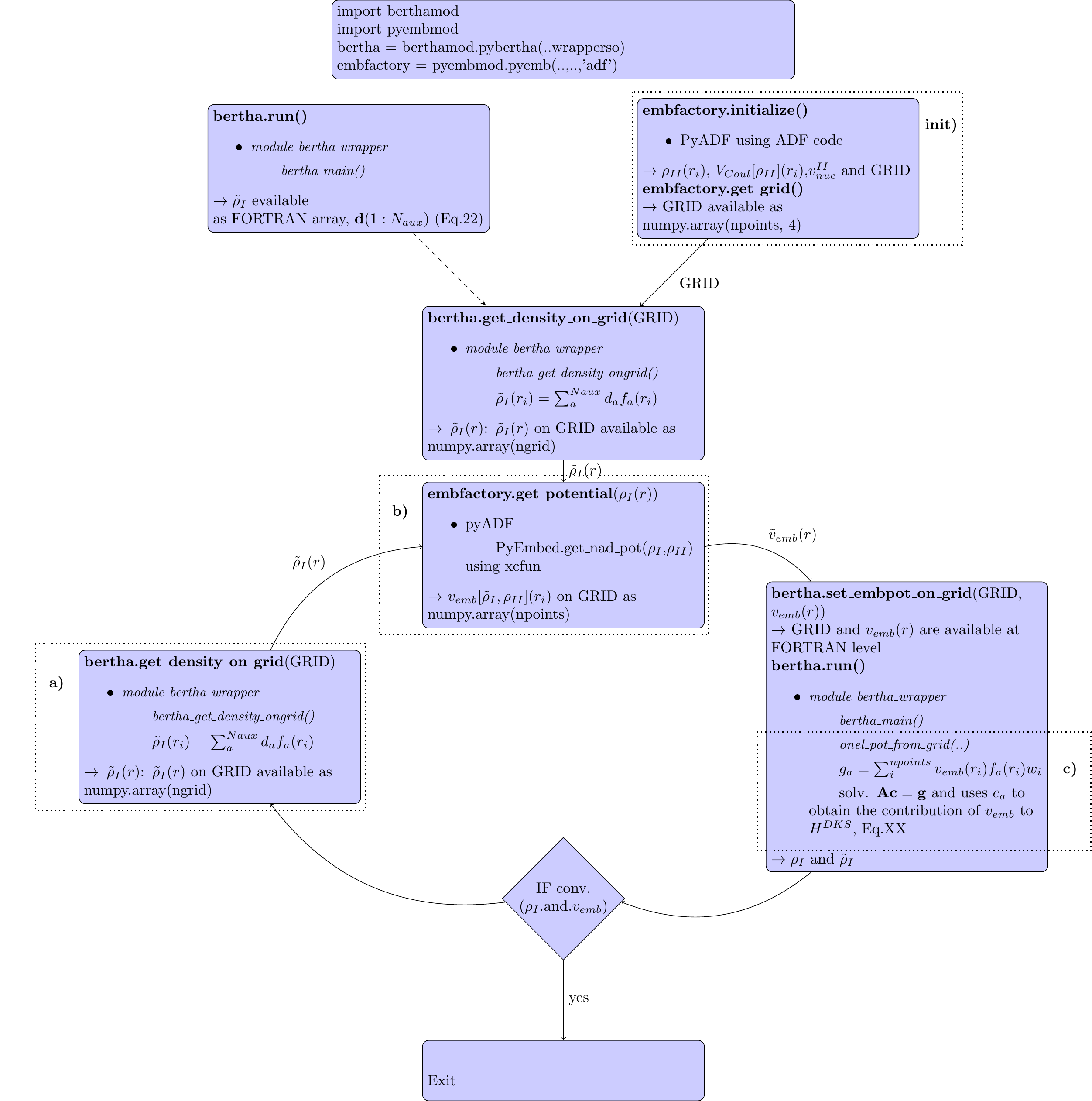}
	\caption{Working flowchart of the Pyberthaemb. In the out-of-loop section the density and electrostatic potential of the environment are obtained as grid functions (init). 
         a) numerical representation of $\tilde \rho(r)$ on the grid
         b) PyEmbed classes are used to calculate the embedding potential.
	 c) projection of the embedding potential onto fitting basis functions.}
        \label{fig:fde_cycle}
\end{figure}

In Figure \ref{fig:fde_cycle} we present the workflow
which emphasizes the interoperability between different tasks and modules or programs
involved including the layers where the actual computations are carried out. 
This schematic picture highlights also how the key quantities, required to implement the DKS-in-DFT FDE
scheme, have different representations along the computation.
As example, we focus on the electron density of the active system, $\tilde \rho_I(r)$.
This quantity is evaluated at the DKS level of theory 
 activated by the \textbf{bertha.run()} method  
within the \textsc{PyBERTHAembed} program. The actual calculation is  done within the FORTRAN  layer.
At this level, $\tilde \rho_I(r)$ is represented in terms of the expansion coefficients of
auxiliary fitting functions ($\bm d$, see Eq.\ref{eq:A_mat}) and is stored  as a FORTRAN array (of dimension $N_{aux}$).
However, this representation is not useful itself for the evaluation of the embedding potential. Indeed, its evaluation and in particular the
non-additive contribution requires that $\tilde \rho_I(r)$ is represented on a grid.
Thus, within \textbf{pyberthaemb.py}, the numerical grid (GRID) evaluated within PyADF (see panel init)
is made available as \textbf{numpy.array}  and via the \textbf{bertha.get\_density\_on\_grid()} method is made
accessible to the FORTRAN layer (\textbf{bertha\_wrapper} module), and stored as a FORTRAN array of dimension $npoints$.
The calculation of the numerical representation of $\tilde \rho_I(r)$ on the grid is done
efficiently in FORTRAN (see in Figure 2, panel a) and the latter is accessible 
within \textbf{pyberthaemb.py} as a \textbf{numpy.array}.
Analogously, different representations are used also for the embedding potential along the workflow.
Note that all quantities accessible from \textbf{pyberthaemb.py}, namely $\tilde \rho_I(r)$, $v^{emb}({r}_k)$ and GRID (labelling the
arrows in figure), are defined as \textbf{numpy.array} that can be easily manipulated within a Python source code.
The computational steps which involve BERTHA are instead implemented in FORTRAN (panel a) and panel c) in Figure)
and have been efficiently parallelized using OpenMP.

\section{Results and Discussion}\label{sec:results}

In the present section we report a series of numerical results mainly devoted
to assess the correctness of our new implementation of DKS-in-DFT FDE scheme. 
In addition we are also reporting the computational cost and 
scalability with respect to the size of both the active and the embedding system.
Finally, we will present an application to a series of Heavy (Rn)
and Super-Heavy elements (Cn, Fl, Og) confined into a C$_{60}$ cage.

\subsection{Initial validation and numerical stability: H$_{2}$O-NH${_3}$}

As already mentioned, 
in this first version of the \textbf{pyembmod} module we include 
the basic procedures previously implemented in the \textsc{Psi4-rt-PyEmbed} code
which are based  on the use of PyADF\cite{Jacob2011a,pyadf-github-v096} and
of the PyEmbed module\cite{pyembed,Schmitt-Monreal2020}.

This poses us in an ideal framework of having a reference non relativistic DFT-in-DFT FDE
implementation where we can have the precise control over all those details, and parameters, from which a
FDE calculation depends on.
Thus, for the sake of a direct comparison,
we selected a simple molecular complex, namely the  H$_2$O-NH$_3$ adduct,  for which the  
relativistic effects are expected to be negligible. For this system we can 
safely compare directly the numerical results of the DKS-in-DFT FDE method, implemented here, 
with respect to those obtained using the DFT-in-DFT FDE scheme in the \textsc{Psi4-rt-PyEmbed} code \cite{DeSantis2020b}.

In the adduct the water molecule is the active
system that is bound to an ammonia molecule, which instead plays the role of
the embedding environment.  In the \textsc{Psi4-rt-PyEmbed} case we use basis
sets obtained by the decontraction of the Gaussian cc-pVDZ, cc-pVTZ and
aug-cc-pVDZ basis sets~\cite{Dunning1989,Kendall1992} for the active
system (these basis sets are referred as cc-pvdz-decon, cc-pvtz-decon
and aug-cc-pvdz-decon, respectively).  The same basis set has been used
in the DKS calculation to define the large component of the G-spinor basis
set.  The corresponding small component was generated using restricted
kinetic balance relation\cite{Dyall1990}.  Noteworthy, for these reference
calculations, we have used an extremely large auxiliary fitting
basis set (A4$_{spdfg}$) which gives an error on the Coulomb energy even
below $10^{-6}$ Eh.  The computational details for the definition of
the environment, including parameters to define the embedding potential,
are identical in both \textsc{PyBERTHAembed} and \textsc{Psi4-rt-PyEmbed}.  In particular,
the basis set used in PyADF for the calculation of the environment frozen
density (ammonia) and the embedding potential is the AUG-TZ2P Slater-type
set from the ADF library\cite{TeVelde2001}. As numerical grid
we used the supermolecular Voronoi Polyhedra  grid defined in ADF 
which is set defining an the integration parameter equal to 4 (this corresponds to a total number of 
33280 grid points). The PBE~\cite{Perdew1998} exchange-correlation functional has been used for the
active system while the  BLYP~\cite{Becke1988,Lee1988} exchange-correlation
functional has been  used for the ammonia molecule.  The Thomas-Fermi and
LDA functionals~\cite{Vosko1980,Slater1951} has been employed for the
non-additive kinetic and non-additive exchange-correlation potential,
respectively.  The molecular structure of the adduct is reported in SI
(Table S.1).  The effect of the environment (ammonia) on the active system
(water) have been evaluated comparing the dipole moment components
and diagonal elements of the polarizability tensor ($\alpha_{xx},
\alpha_{yy}$ and $\alpha_{zz}$) of the isolated (Free) respect to the embedded
(Emb) water. We note here that the quoted polarizability values are not those for the supermolecular system, but only for the active subsystem.

The numerical results, reported in Table~\ref{tab:vk},
show an evident quantitative agreement between the
two implementations. Indeed, both the variations
induced by the presence of the embedding system ($\Delta$ values)
and the absolute values show a good agreement.
Noteworthy, independently by the basis set used, the differences are
below of 0.001 a.u. and 0.01 a.u.  for the dipole moment components
and for the polarizability tensor components, respectively.  We mention
that we also performed the calculations increasing the speed of light
by 1 order of magnitude (i.e., c = 1370.36 a.u.) to approximate the
non-relativistic limit and, as expected, we obtain almost
indistinguishable results (see Table S.2).  All the above findings make us
confident that our implementation is both numerically stable and correct.

\begin{table}
\begin{tabular}{rrrrrrrrr}
       & \multicolumn{3}{c}{\textsc{Psi4-rt-PyEmbed}}    &  &   \multicolumn{3}{c}{\textsc{PyBERTHAembed}}                                 \\
       \cline{2-4}                                        \cline{6-8}   \\
               & Free     & Emb    &$\Delta$  &  & Free      & Emb      & $\Delta$               \\
	& \multicolumn{7}{c}{}                             \\
	& \multicolumn{7}{c}{a) aug-cc-pvdz-decon}                             \\
	& \multicolumn{7}{c}{}                             \\
$\mu_x$        &-0.35403  & -0.49351 & -0.13948 &  &   -0.35328        &  -0.49279 & -0.13951  \\    
$\mu_y$        &-0.62058  & -0.62976 & -0.00918 &  &   -0.61908        &  -0.62812 & -0.00904  \\
$\mu_z$        &-0.00025  & -0.00026 & -0.00001 &  &   -0.00025        &  -0.00027 & -0.00002  \\
$|\mu|$        & 0.71446  &  0.80009 &  0.08563 &  &    0.71279        &   0.79836 & 0.08557   \\
$\alpha_{xx}$  &10.34     &  9.79    &  -0.55   &  &   10.36           &  9.80     & -0.56     \\
$\alpha_{yy}$  &9.91      & 10.26    &   0.35   &  &   9.92            &  10.27    &  0.35     \\
$\alpha_{zz}$  &9.62      & 10.16    &   0.54   &  &   9.64            &  10.18    &  0.54     \\
$\alpha_{iso}$ &9.96      & 10.07    &   0.11   &  &   9.97            &  10.08    &  0.11     \\
	& \multicolumn{7}{c}{}                             \\
	& \multicolumn{7}{c}{b) cc-pvdz-decon}                             \\
	& \multicolumn{7}{c}{}                             \\
$\mu_x$        & -0.38085& -0.50083  & -0.11998 &  &  -0.37992       & -0.50006 & -0.12014 \\      
$\mu_y$        & -0.67072& -0.67187  & -0.00115 &  &  -0.66911       & -0.67023 & -0.00112 \\
$\mu_z$        & -0.00027& -0.00031  & -0.00004 &  &  -0.00027       & -0.00031 & -0.00004 \\
$|\mu|$        & 0.77130 &  0.83800  & 0.06670  &  &   0.76944       &  0.83622 &  0.06678 \\
$\alpha_{xx}$  & 7.35    & 6.79      & -0.56    &  &   7.36          & 6.80     & -0.56    \\
$\alpha_{yy}$  & 6.27    & 6.38      &  0.11    &  &   6.28          & 6.39     & 0.11     \\
$\alpha_{zz}$  & 3.70    & 3.71      &  0.01    &  &   3.70          & 3.71     & 0.01     \\
$\alpha_{iso}$  & 5.77    & 5.63      & -0.14    &  &   5.78          & 5.63     & -0.15    \\
	& \multicolumn{7}{c}{}                             \\
	& \multicolumn{7}{c}{c) cc-pvtz-decon}                             \\
	& \multicolumn{7}{c}{}                             \\
$\mu_x$        &  -0.36464& -0.49344 & -0.12880 &  & -0.36377 & -0.49267 & -0.12890\\
$\mu_y$        &  -0.64037& -0.64583 & -0.00546 &  & -0.63883 & -0.64430 & -0.00547\\
$\mu_z$        &  -0.00025& -0.00028 & -0.00003 &  & -0.00026 & -0.00028 & -0.00002\\
$|\mu|$        &  0.73691 &  0.81276 &  0.07585 &  &  0.73514 &  0.81108 &  0.07593\\
$\alpha_{xx}$  &  8.52    &  7.95    & -0.57    &  & 8.53     & 7.96     & -0.57  \\
$\alpha_{yy}$  &  7.80    &  7.94    & 0.14     &  & 7.81     & 7.95     & 0.14   \\
$\alpha_{zz}$  &  5.77    &  5.80    & 0.03     &  & 5.78     & 5.82     & 0.04   \\
$\alpha_{iso}$ &  7.36    &  7.23    & -0.13    &  & 7.37     & 7.24     & -0.13  \\
\hline
\end{tabular}
	\caption{Dipole moment (components $\mu_x$, $\mu_y$, $\mu_z$ and
	module $|\mu|$) and dipole polarizability (tensor diagonal components
	$\alpha_{xx}$, $\alpha_{yy}$, $\alpha_{zz}$ and isotropic
	contribution $\alpha_{iso}$) of both the isolated (Free)
	and embedded (Emb) water molecule. In the embedded water
	molecule, an ammonia molecule is used as environment. Data
	have been obtained using our new \textsc{PyBERTHAembed}  implementation
	and the reference \textsc{Psi4-rt-PyEmbed} implementation (see text for
	details). The shift $\Delta$ is also reported.	All numerical
	data are reported in atomic units (a.u.). The diagonal components
	of the dipole polarizability tensor have been calculated using
	a finite field approach using an external electric of 0.001 a.u.
	}
\label{tab:vk} \end{table}

\begin{table}[h!]
\begin{tabular}{ccccccc}
	&  A2$_{s}$          & A2$_{sp}$          & A2$_{spd}$  &  A2$_{spdfg}$ & A3$_{spdfg}$ & A4$_{spdfg}$  \\
     N$_{aux}$ &   (19)         &  (67)            &  (163)  &   (338)   & (403)    &  (544)  \\
\hline                                                                                        
	$\mu_x$ & -0.49845& -0.50555&-0.49264& -0.49250& -0.49261& -0.49267\\
	$\mu_y$ & -0.65654& -0.64784&-0.64804& -0.64415& -0.64445& -0.64429\\
	$\mu_z$ & -0.00034& -0.00051&-0.00024& -0.00028& -0.00028& -0.00028\\
	$|\mu|$ & 0.824322&  0.82175& 0.81404&  0.81086&  0.81116&  0.81107\\
$\alpha_{xx}$   & 8.46               & 7.95                  & 7.96          & 7.96              & 7.96           & 7.96          \\
$\alpha_{yy}$   & 7.19               & 8.12                  & 7.95          & 7.95              & 7.95           & 7.95          \\
$\alpha_{zz}$   & 3.90               & 6.08                  & 5.80          & 5.81              & 5.81           & 5.82          \\
$\alpha_{so}$   & 6.52               & 7.38                  & 7.24          & 7.24              & 7.24           & 7.24         \\
$\Delta{E}_{J}$ & 1.51(10-2) & 4.47(10-3)            & 2.8(10-4)     & 4.8(10-6)         & 1.9(10-6)      &5.0(10-7)   \\
\hline
\end{tabular}
\caption{
	Dipole moment (components $\mu_x$, $\mu_y$, $\mu_z$ and
        module $|\mu|$) and dipole polarizability (tensor diagonal components
        $\alpha_{xx}$, $\alpha_{yy}$, $\alpha_{zz}$ and isotropic
	contribution $\alpha_{iso}$) of the embedded water molecule (water-ammonia system). 
	Data have been obtained with our new \textsc{PyBERTHAembed} implementation (using a G-spinor basis functions derived 
	from  the cc-pvtz-decon basis) 
	and several auxiliary density fitting basis sets (A2$_{s}$, A2$_{sp}$, A2$_{spd}$, A2$_{spdfg}$, A3$_{spdfg}$ and A4$_{spdfg}$).
	The size of different fitting basis sets (N$_{aux}$) are reported in parenthesis.
	$\Delta{E}_J$ is the absolute error on the Coulomb energy due to the density fitting.
        The diagonal components of the dipole polarizability tensor have been calculated using
        a finite field approach using an external electric of 0.001.
	All numerical data are reported in atomic units (a.u.).
	See text for the fitting basis set definition and further details.
        }
\label{tab:fitting}
\end{table}

As we have extensively described in the previous section, our implementation 
strongly benefits of the use of auxiliary
fitting functions, both in the definition
of the  embedding potential and
as intermediate quantities to obtain the G-spinor matrix representation of the embedding potential. Thus,
it appears mandatory to investigate the impact 
of the quality of the density fitting
basis set on the final results of DKS-in-DFT  FDE calculations.
In addition to the limit auxiliary fitting basis set 
employed above
we generated five fitting basis sets
(A2$_{s}$, A2$_{sp}$, A2$_{spd}$, A2$_{spdfg}$ and A3$_{spdfg}$) of
increasing accuracy. We have adopted a procedure which is strictly related
with that proposed by K\"{o}ster et al. and employed in Demon2K code (see appendix of
Ref.\citenum{Calaminici2007}). All the fitting basis sets are explicitly reported
in SI, while the results are reported in
Table \ref{tab:fitting}. In the Table we also show the absolute error in the Coulomb
energy ($\Delta{E}_{J}$), which is
the quantity that is variationally optimized in the fitting procedure and
typically regarded as its quality index.
This numerical test shows that the use of density fitting does not introduce any
significant instability in the DKS calculation of the active system, also in
presence of the embedding potential. 
The $\Delta{E}_{J}$ values are showing a convergent trend of both the dipole moment components and the polarizability when the quality of the fitting basis set is increased.
The fitting basis sets A2$_{s+}$
and A2$_{sp+}$, bearing only s- and p-type Hermite Gaussian functions, have 
values of $\Delta{E}_{J}$ larger than 1 mEh and 
are clearly inadequate to reproduce the reference results.
Very accurate results can already be obtained starting from the A2$_{spd}$ auxiliary basis
set (i.e., 163 functions for the water molecule).
It is interesting to note that the $\Delta{E}_{J}$ associated with this basis set 
is of the same order of magnitude of that typically required (0.1 mEh per
atom) in standard calculations based on density fitting without including FDE.
Thus, these preliminary results suggest that the variational density fitting 
scheme can safely be applied in the implementation  of DKS-in-DFT method 
without jeopardizing its accuracy.

\subsection{Computational efficiency : gold clusters in water}

It is interesting now to put forward some assessments on the
computational efficiency of our DKS-in-DFT FDE implementation, together with 
its scaling properties in terms of time statistics and memory usage. 
This analysis will give us a detailed overview of the
the computational burden, and possible bottlenecks, along the relatively complex 
workflow we implemented (using different quantum chemistry packages and programming
languages). Furthermore, it will be a solid starting
point for future optimizations and developments (e.g. DKS-in-DKS or coupled 
real time DKS-in-DKS). As a test case we have chosen a series
of gold clusters (Au$_2$, Au$_4$, Au$_8$) embedded using  an increasing
number (5, 10, 20, 40 and 80) of water molecules. In all cases, for Au the
large component of the G-spinor basis set was generated by uncontracting
double-$\zeta$ quality Dyall's basis sets ~\cite{dyall2002relativistic, dyall2006relativistic,dyall2010revised}
augmented with the related polarization and correlating functions
($24s19p12d9f1g$),  while the corresponding small component basis was
generated using the restricted kinetic balance relation.  For the water
molecules of the environment we used the DZ Slater-type set from the ADF
library\cite{TeVelde2001}.  The supermolecular grid defined in PyADF, 
corresponding to an integration parameter of 4 in the ADF package,
has been used.
The  BLYP~\cite{Becke1988,Lee1988} exchange-correlation functional
is used for the ground state calculation of the embedding system,
while the Thomas-Fermi and LDA functionals~\cite{Vosko1980,Slater1951}
have been employed for the non-additive kinetic and non-additive
exchange-correlation potential, respectively. The molecular
structure of all the adducts are available in 
Ref.\citenum{matteo_de_santis_2022_6343894}.  All the
calculations have been performed on a Dual Intel(R) Xeon(R) CPU E5-2684
v4 running at 2.10GHz, equipped with 251 GiB of RAM. We used the Intel
Parallel Studio XE 2018 \cite{blair2012parallel} to compile the FORTRAN
code and Python 3.8.5 (from Anaconda, Inc.) and  NumPy version 1.19.2
for the Python code.  We used PyADF\cite{Jacob2011a,pyadf-github-v096} as
recently ported to Python3\cite{pyadf-github-v096-python3}, ADF(version
2019.307) for the core DFT calculations of the environment and {\it XCFun}
library
(version 1.99).\cite{Ekstrom2010,xcfun:2019,xcfun-python3}

The results are reported in Table \ref{tab:time_stats} and
in Table \ref{tab:time_stats1}, where, together with the total elapsed time (t$^d$)
for each SCF iteration including the FDE contribution,
we also partition between different tasks related with the FDE implementation,
namely:   a) numerical representation of active system fitted density on grid,  $\tilde \rho(r)$;
b) calculation of the non-additive terms of embedding potential by PyADF (with the PyEmbed class);
c) projection of the embedding potential onto fitting basis functions.
In the Tables
we also report the maximum memory usage for the SCF procedure ("Mem"), the number of 
points of grid and the timing for the "init" phase 
which involves: the evaluation of the ground state electronic density of the 
environmental together with the associated 
Coulomb potential, and their mapping on the numerical grid. 
We recall that the electron density of the environment is kept frozen, thus
this initial step is done once at the beginning of the procedure.
All  tasks are also highlighted (using the same labeling: a, b, c and init) 
in Figure \ref{fig:fde_cycle}.

As general remark we may state that the FDE contribution to the total time is relatively 
small. 
By increasing the size of the active system (Au$_2$, Au$_4$ and Au$_8$),
and keeping fixed the environment (using ten water molecules), see Table 3,
the relative impact of the FDE computational phase decreases.  
It passes from  13.3\% for Au$_2$(H$_2$O)$_{10}$ to 0.9\% for Au$_8$(H$_2$O)$_{10}$.
This may be expected since tasks a), b) and c) 
have a more favorable scaling than the DKS calculation (i.e., $O(N^{3})$).
The computational
cost for the step a) and c) is proportional to the product $N^{aux} \cdot N_{grid points}$, where: $N^{aux}$ is  the total number of the auxiliary
fitting functions in the active system, and $N_{grid points}$ total number of grid points. 
Thus, the computational cost should scale as $O(N^{2})$ (being $N$ the dimension of the active system).
The actual scaling is much lower (i.e., slightly higher than $O(N)$) 
mainly because 
%we are far from  the limit case, and 
the total 
grid points are largely dominated by the environmental system (see number of grid points, $N_{grid points}$, as reported in Table \ref{tab:time_stats} ). 
Concerning the step b) and considering the fact that the environment is maintained fixed,
it scales, as expected,  linearly with number of points of the grid, $N_{grid points}$. 
The maximum use of memory, 
during the whole DKS-in-DFT FDE procedure, increases with respect to the number of Au atoms being
N$^{1.7}$, which is close to the theoretical value N$^{2}$.

When we fix the active system (Au$_4$) 
increasing instead the size of  the environment, see
Table \ref{tab:time_stats1}, 
the relative computational cost to include the embedding 
passes from 2.2\%, in the case of Au$_4$@(H$_2$O)$_5$,
to 16.1\% for Au$_4$@(H$_2$O)$_{80}$. In this case,  all tasks associated with the FDE procedure (a,b,
and c) have a computational burden which increases linearly with the size 
of the environment (and the number of total grid points, see Figure S1 in SI),
while the  maximum memory usage during the SCF procedure is almost  independent
from the number of water molecules in environment, as only a slight increase can be observed.
\begin{table}
	\caption{Elapsed real time (s). $^a$: fitted density on grid; $^b$ : calculation of the non-additive terms of embedding potential by PyADF (with PyEmbed classes);
        $^c$ : projection of the embedding potential onto fitting basis functions $^d$ : Total time for a single DKS self-consistent field interaction. 
	All the calculations have been performed in a Dual Intel(R) Xeon(R) CPU E5-2684 v4 running at 2.10GHz,
equipped with 251 GiB of RAM. We used the Intel Parallel Studio XE 2018 \cite{blair2012parallel} to compile the
FORTRAN code and instead Python 3.8.5 Anaconda. Inc and  NumPy version 1.19.2 for the Python code. See text for further details.}
\label{tab:time_stats}
	\begin{tabular}{llllllll}
\hline
         System		        & t$^a$&t$^b$  & t$^c$ & t$^d$    & Mem(MB) & grid points & init embfactory  \\
\hline
	Au$_2$(H$_2$O)$_{10}$   & 1.47 & 2.74  & 1.48  &  42.68   & 1165    & 213248      &  138.9 (8.8)    \\
	Au$_4$(H$_2$O)$_{10}$   & 3.06 & 2.84  & 3.09  &  260.16  & 2164    & 221824      &  127.9 (9.0)    \\
	Au$_8$(H$_2$O)$_{10}$   & 6.62 & 3.05  & 6.70  &  1849.90 & 7572    & 237824      &  154.7 (9.8)    \\
\end{tabular}
\end{table}

\begin{table}
        \caption{
Elapsed real time (s).
$^a$: fitted density on grid; $^b$ : calculation of the non-additive terms of embedding potential by PyADF (with PyEmbed classes);
        $^c$ : projection of the embedding potential onto fitting basis functions $^d$ : Total time for a single DKS self-consistent field interaction.
        All the calculations have been performed in a Dual Intel(R) Xeon(R) CPU E5-2684 v4 running at 2.10GHz,
equipped with 251 GiB of RAM. We used the Intel Parallel Studio XE 2018 \cite{blair2012parallel} to compile the
	FORTRAN code and instead Python 3.8.5 Anaconda. Inc and  NumPy version 1.19.2 for the Python code. See text for further details.}
\label{tab:time_stats1}
\begin{tabular}{lllllllll}
\hline
        System                        & t$^a$ &t$^b$  & t$^c$ &  t$^d$    & Mem(Mb) & grid points & init embfactory  \\
\hline
	Au$_4$(H$_2$O)$_5$      & 1.97  & 1.84  & 1.99  &  257.70   & 2137    & 143232      &  102.3 (5.9)    \\
	Au$_4$(H$_2$O)$_{10}$   & 3.06  & 2.84  & 3.09  &  260.16   & 2164    & 221824      &  127.9 (9.0)    \\
	Au$_4$(H$_2$O)$_{20}$   & 5.04  & 4.71  & 5.07  &  260.44   & 2225    & 366336      &  215.4 (14.8)   \\
	Au$_4$(H$_2$O)$_{40}$   & 8.69  & 8.09  & 8.71  &  270.19   & 2331    & 630144      &  641.0 (25.8)   \\
	Au$_4$(H$_2$O)$_{80}$   & 16.27 & 15.19 & 16.40 &  295.65   & 2354    &1184896      &  2600.1 (47.8)  \\
\end{tabular}
\end{table}

As already mentioned in the previous sections we have recently developed an OpenMP
parallel version of BERTHA which can be easily used directly via the Python API \cite{Belpassi2020}. This only requires
the \textbf{berthamod} module, which refers to the shared object \textbf{libberthaserial.so}, to be compiled with OpenMP flag set. Thus, here we have
extended the OpenMP parallelization to those steps of the FDE procedure in which the BERTHA code 
is directly involved, namely the 
steps a) and c), see above. The results are given for the Au$_4$(H$_2$O)$_{80}$ system and are 
reported in Table \ref{tab:time_parallel}.
These  steps have been efficiently parallelized and 
we are able to achieve a speed-up of 31.1 and 29.8 using 32 threads, respectively for step a and d.
Noteworthy, for this parallel implementation the FDE phase is
about 45\% of the total elapsed time and is dominated by the computation task that remains serial part.
Indeed, using 32 threads the task b takes 15.20 sec of the total 35.8 sec necessary for  each  SCF iteration. 
This task, that is related to the generation of the non-additive kinetic and exchange-correlation 
on grid is currently carried out by the PyEmbed component in PyADF.
Regarding the  memory usage, in our OpenMP implementation we observe a linear growth of memory usage
 with respect to the number of
the employed threads. This is somehow expected due to the obvious data replication in the OpenMP implementation. 
Despite one may expect that there may be room for a further optimization, we
note that  even in the current version the implementation  is not memory-bound.
In the case of 32 threads we found a maximum memory usage of about
11 GiB which demonstrates that such kind of calculations, and  even larger ones, can be routinely 
carried out on the current multi-core architectures which may easily 
achieve 64 to 128 cores and 512 to 1024 GiB per node.

\begin{table}
        \caption{Elapsed real time (s) for the Au$_4$(H$_2$O)$_{80}$ system. 
$^a$: fitted density on grid; $^b$ : calculation of the non-additive terms of embedding potential by PyADF (with PyEmbed classes);
        $^c$ : projection of the embedding potential onto fitting basis functions $^d$ : Total time for a single DKS self-consistent field interaction.
        All the calculations have been performed in a Dual Intel(R) Xeon(R) CPU E5-2684 v4 running at 2.10GHz,
equipped with 251 GiB of RAM and all the running time have been obtained using the dynamic schedule in OpenMP. 
	We used the Intel Parallel Studio XE 2018 \cite{blair2012parallel} to compile the
FORTRAN code and instead Python 3.8.5 Anaconda. Inc and  NumPy version 1.19.2 for the Python code. See text for further details.}
\label{tab:time_parallel}
\begin{tabular}{llllllll}
\hline
n. threads & t$^a$  &t$^b$    & t$^c$  & t$^d$  &   Mem(Mb) & init embfactory  \\
  1 & 16.16 & 15.16  & 16.40   & 291.06 & 2343       &  2634.4 (48.5)  \\
  2 & 8.12  & 15.23  & 8.30    &  158.20& 2638       &  2633.1 (48.8)   \\
  4 & 4.07  & 15.23  & 4.11    &  91.10 & 3210       &  2631.4 (48.5)   \\
  8 & 2.03  & 15.32  & 2.06    &  60.48 & 4377       &  2655.7 (48.4)   \\
  16& 1.04  & 15.23  & 1.10    &  43.67 & 6632       &  2603.4 (48.5)   \\
  32& 0.52  & 15.20  & 0.55    &  35.80 & 11020      &  2601.4 (49.1)   \\
\hline
\end{tabular}
\end{table}

\subsection{The generation of atom-endohedral fullerenes model potentials}

We conclude our work by showcasing how we can leverage our FDE implementation to determine fullerene-atom model potentials, that are applicable for species across the periodic table.

Over the past decades, it has been recognized that fullerenes can serve as containers for other,
smaller species\cite{popov2013endohedral}. As such, there has been considerable interest in
understanding how such smaller species behave under confinement, both from a fundamental point of
view as well as due to possible technological applications we mention: the potential use as seed
materials in solid state quantum computation \cite{ju2007two}, and the use as agents for
improving the super-conducting ability of materials \cite{takeda2006superconductivity}. 

From a more fundamental perspective, the study of how atomic species behave under such
confinement is a particularly active domain.  With respect to the use of theoretical approaches, a number of studies have been reported that employed, in most cases, simple models of the C$_{60}$ cage potential to represent the confinement potential~\cite{dolmatov2010confinement, amusia2005dramatic, lyras2005electronic}. 
%Evidently, a clear limitation is the omission of dynamical coupling of the electrons I guess in the case of 
%photoionizzation study and indeed \ciet{madjet2007giant},
A model where only electrons of the guest atoms are considered while  
the $C_{60}$ cage is modelled, in most cases, by a short-range attractive $V_c(r)$  spherical potential defined as follows:

\begin{equation}
V_{c}(r) = 
\left\{\begin{matrix}
U_{0}, & if \ \ r_{0}  \leq  r \leq r_{0} + \Delta  \\ 
0, & otherwise
\end{matrix}\right.
\label{eq:modelpot}
\end{equation}

where $r_0 = 5.8 \ a.u.$ and $U_{0} = -0.30134 \ a.u. $ and $ \Delta = 1.9 \ a.u$ represent
the finite thickness of the spherical potential\cite{dolmatov2010confinement}. Other model potentials have been proposed \cite{baltenkov1999resonances}, as well as other approaches to avoid numerical instability related to the sharp form of those potentials  \cite{madjet2008photoionization,nascimento2010study}.
An alternative approach may be to start from the embedding potential generated in the FDE scheme,
which is expected to be highly accurate and without artificial discontinuities. In the following, we propose a possibly general procedure to build 
model potentials for atomic calculations and, with this aim, 
we compare the results, in terms of HOMO-LUMO gap, for a set of heavy atoms,
obtained using the Frozen Density Embedding (FDE) procedure respect to the simple 
spherical potential (SPM) modelled by Eq. \ref{eq:modelpot}. Practically we applied the FDE scheme to a set of neutral endohedral fullerenes
 A@C$_{60}$ (A=Rn,Og,Fl,Cn), where the atom A (i.e., active system) is embedded in a fullerene (i.e., environment) and always placed at the exact center of the C$_{60}$.
Finally, by comparing 
the embedding potential (EMBP) and its spherical average with respect to the cited spherical potential model (SPM), we 
propose a simple numerical recipe that  can be used within the FDE scheme to possibly extract more 
accurate potentials to be tested in atomic calculations.

Before proceeding in the comparison of different models, we have compared for the Rn atom the ability of FDE to capture environment effects on orbital energies, with respect to standard (supramolecular) DFT calculations. Our results, shown in Figure~S2 in the supplementary information, show that there is a good agreement between the FDE and supramolecular calculations, with FDE yielding overall slightly larger orbital energy shifts, that are nevertheless very homogeneous across the different orbitals. From these results we conclude that FDE is, in effect, capable of correctly describing the fullerene cage's effect onto the atom's electronic structure.

\begin{figure}[h!]
\centering
\includegraphics[scale=0.4]{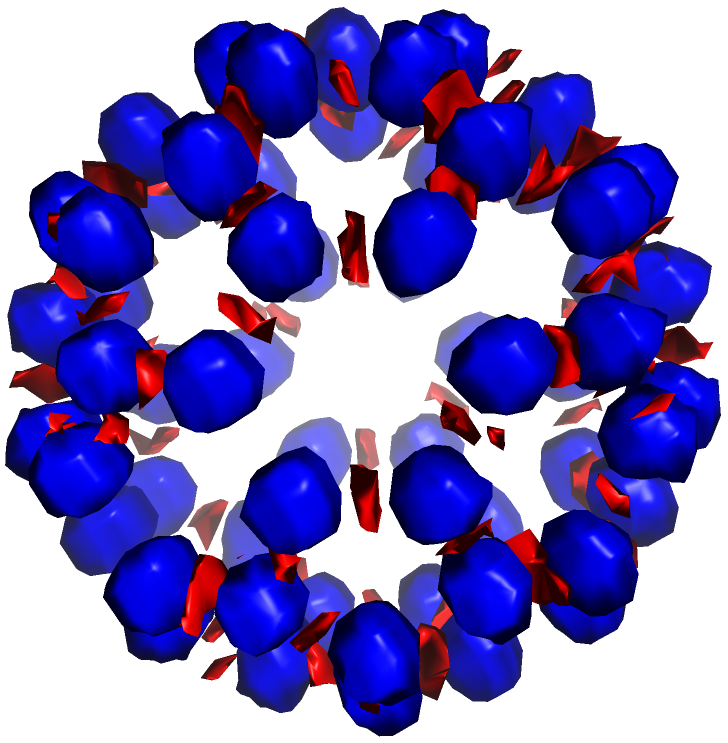}
\caption{The embedding potential (EMBP) in blue (negative) and red (positive), computed for the Rn@C$_{60}$ system.
We report the contour plot at $ \pm 0.3$ a.u. It is important to underlined as the plotted values are the result
of a Nearest-neighbor interpolation performed starting from the original non homogeneous ADF grid.}
\label{fig:c60fields}
\end{figure}

All calculations, reported in the following, were carried out using a basis set for the active system (i.e., A=Rn,Og,Fl,Cn) generated by uncontracting triple--$\zeta$ quality Dyall's basis sets 
\cite{Dyall2004,dyall2010revised,dyall2006relativistic,dyall09_12638}
augmented with the related polarization and correlating functions. Final basis set schemes are as follows: Cn (32s29p20d14f7g2h), 
Rn (31s27p18d12f4g1h), Fl and Og (31s30p21d14f6g2h).

For all the elements we used auxiliary basis sets already employed in Ref.\citenum{Rampino2015}
and are explicitly reported in SI.
While for the
environment (i.e., the C$_{60}$), computed using the ADF code, we use the TZP basis set. In both cases we use the BLYP \cite{Becke1988,Lee1988} exchange-correlation functional, while for the nonadditive kinetic and
nonadditive exchange-correlation terms in the generation of the
embedding potential, the Thomas-Fermi and LDA functionals
are used, respectively.

\begin{figure}[h!]
\centering
\includegraphics[scale=0.32]{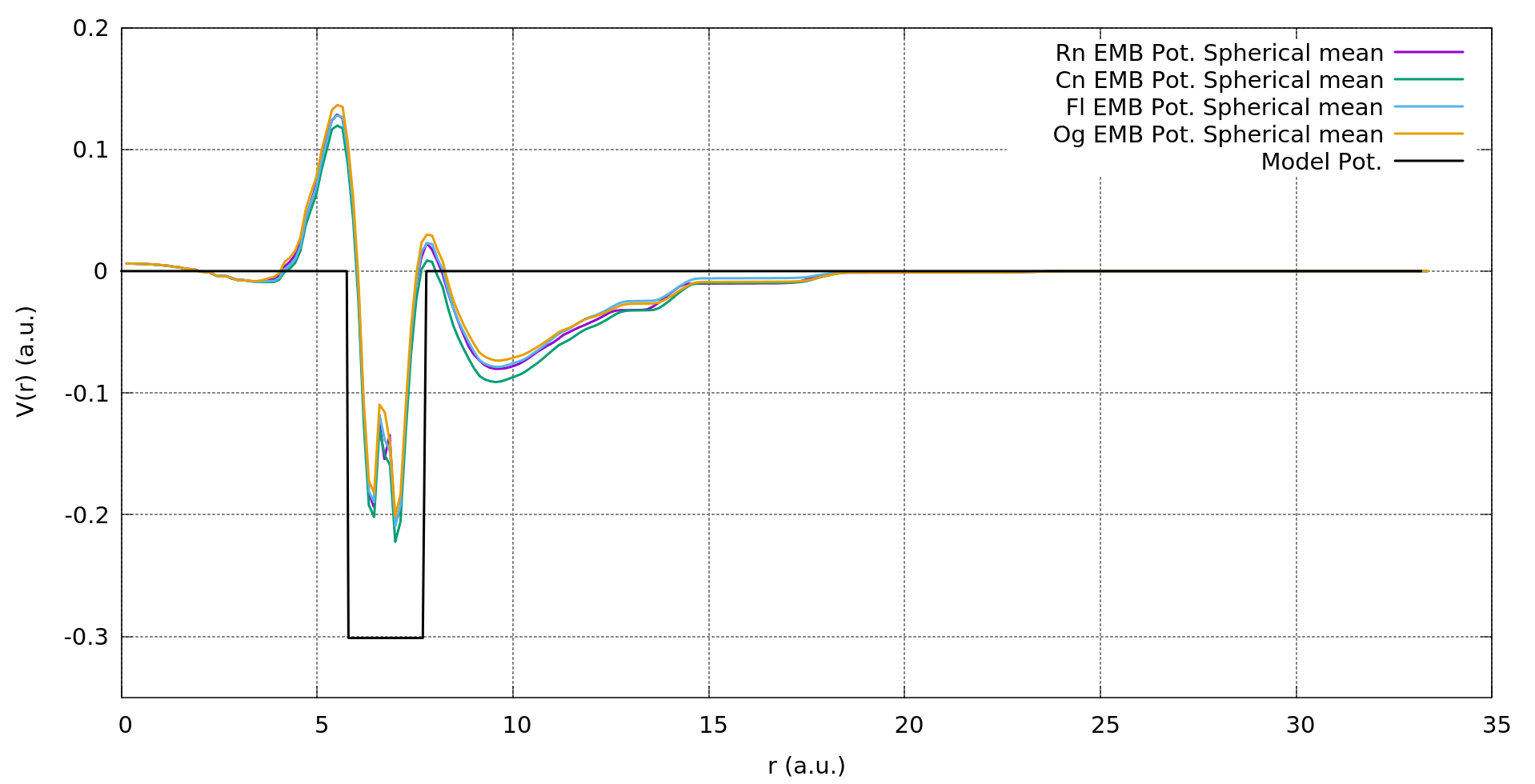}
\caption{The spherical average of the Rn, Og, Cn and Fl embedding potential together with the simple short-range 
attractive spherical potential as reported in Eq. \ref{eq:modelpot}.}
\label{fig:modelpots}
\end{figure}

Figure \ref{fig:c60fields} reports the EMBP of the Rn@C$_{60}$ system. 
The EMBP shows positive values centered at nuclei positions, and negative values 
located in correspondence of the bonds. 
As one may expect the EMBP potential, while maintaining an overall spherical shape, is clearly 
different with  respect to a simple short-range  attractive $V_c(r)$  spherical potential. Indeed, if we consider 
the spherical average of the EMBP (see SI for details on the spherical average procedure employed) extracted for the various A@C$_{60}$ systems, as reported in Figure \ref{fig:modelpots}, while
the EMBP seems to detect the same short-range attractive values surely it shows a more complex
radial structure. 
The spherical average of the EMBP shows a positive repulsive value immediately before the inner C$_{60}$ surface and, maybe more importantly, never completely goes to zero, not even at the center of 
the fullerene where the atom A is placed.
\begin{table}[]
\begin{tabular}{lllll}
\hline
     Atom     & SPM & FDE C$_{60}$ & EMBP              & Rn based EMBP \\ 
              & &              & Spherical Average & Spherical Average\\
\hline                  
   Rn         & 0.118680 & 0.209643  &  0.207990  & \textellipsis \\ 
   Cn         & 0.055807 & 0.144685  &  0.144693  & 0.147514 \\
   Fl         & 0.072251 & 0.110471  &  0.110467  & 0.108101 \\
   Og         & 0.148040 & 0.139842  &  0.139300  & 0.134420 \\
\end{tabular}
	\caption{HOMO-LUMO gap energies (a.u.).}
\label{tab:gaps}
\end{table}

\begin{figure}[h!]
\centering
\includegraphics[scale=0.32]{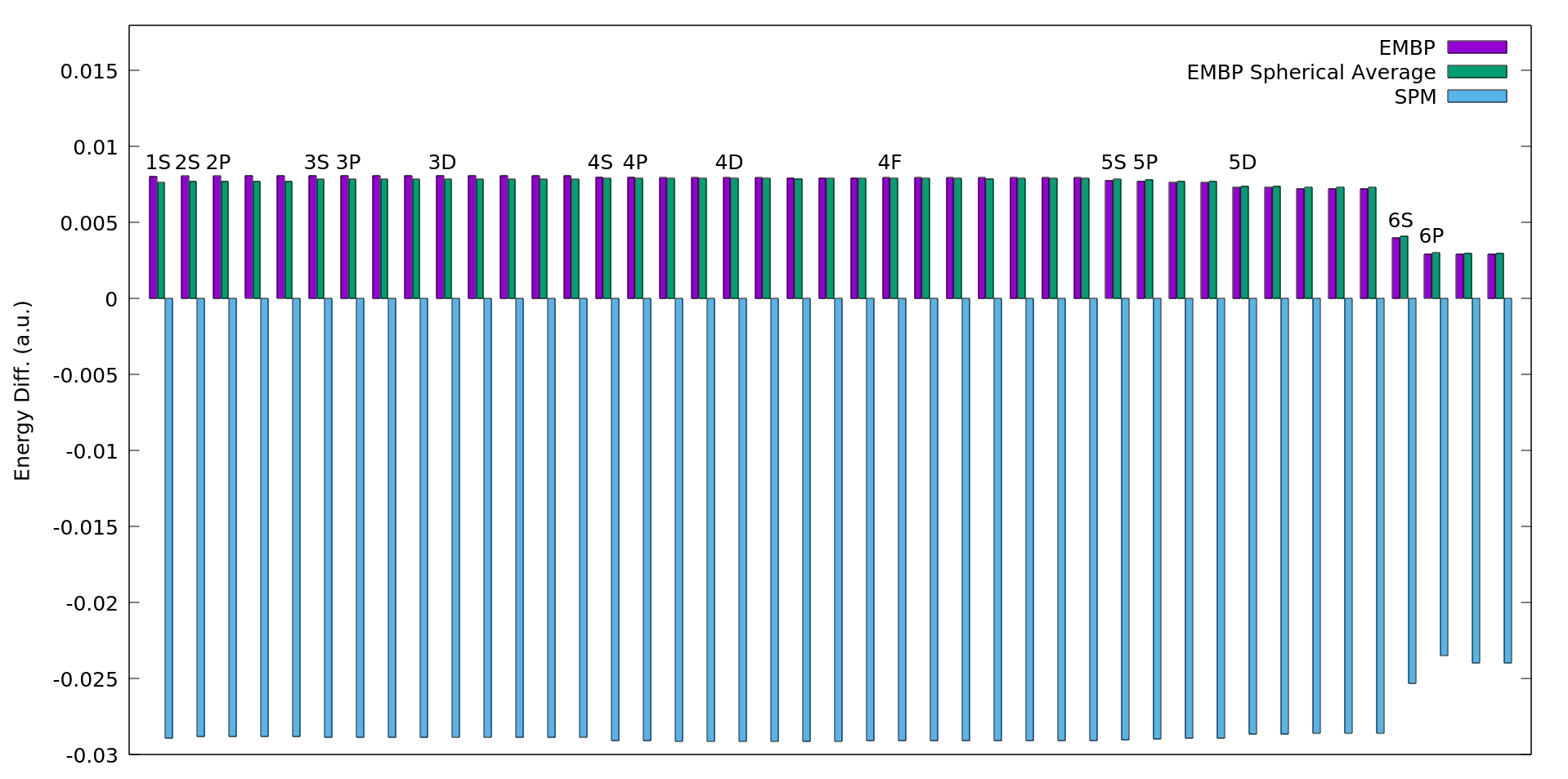}
\caption{Differences in orbital energies with respect to the isolated Rn atom for all occupied orbitals for: the SPM model, the EMBP and the Spherical average of the EMBP.}
\label{fig:differences}
\end{figure}

Not surprisingly, the final results (see Table \ref{tab:gaps}) in terms of HOMO-LUMO gap for the 
spherical model potential are quite different respect to the results
obtained using the FDE procedure. Indeed, as we mentioned, the overall shape of the EMBP is quite 
different respect to a simple spherical ones, see Figure \ref{fig:c60fields}. Nevertheless, is interesting to note as
the spherical average seems to work well. Comparing the results obtained from the full FDE procedure respect to the 
ones computed using a model potential that is the spherical average of the EMBP, respectively columns 3 and 4 of Table \ref{tab:gaps},  
one can easily note that the spherical average is able to well reproduce the electronic structures of the active system (i.e., the central atom) including  HOMO-LUMO gaps values with an error that is generally less the 1\%.
Similar conclusions can be drawn looking at Figure \ref{fig:differences}, where we report instead all the differences in orbital energies with respect to the isolated Rn atom for all the occupied orbitals. Once again both the EMBP (i.e., the FDE procedure) and its spherical average  lead to similar results. Instead using the simple spherical model the energy shift is always the opposite.     

Finally, and maybe more interestingly, if we consider column 5 of Table \ref{tab:gaps} we see that
we can obtain the HOMO-LUMO gaps for all the atoms with an error that is always less then 4\% using a Rn based EMBP spherical average. 
Thus, we computed the spherical average of the EMBP for the Rn@C$_{60}$ system and using this single model potential we have been able to 
quantitatively well reproduce the HOMO-LUMO gap for all the other A@C$_{60}$ systems. 
This latter result let us clearly envision a practical approach to be used to build model potential as a result of the FDE procedure.

\section{Conclusions and perspectives}

Including environmental effects based on first-principles is of
paramount importance in order to obtain an accurate description of
molecular species in solution and in confined spaces.  Among others,
the Frozen Density Embedding (FDE) density functional theory 
represents a embedding scheme in which
environmental effects are included by considering explicitly the environmental
system by means of its "frozen" electron density.  In the present paper,
we reported our extension of the full 4-component relativistic Dirac-Kohn-Sham method, as
implemented in the BERTHA code, 
to include environmental and confined effects with the FDE
scheme (DKS-in-DFT FDE) using the PyADF framework. 
We described how its complex workflow associated
with its implementation can be enormously facilitated by the fact that both BERTHA (PyBERTHA) and
PyADF, with their Python API, they gave us an ideal framework of development.
The recent development of the \textsc{Psi4-rt-Embed} code\cite{DeSantis2020b}, which is also based on  
PyADF for FDE while uses Psi4Numpy code for the active system, represented an ideal reference
implementation to assesses the correctness of our new DKS-in-DFT FDE implementation.

\textsc{PyBERTHAembed} uses the density
fitting technique at the key points of the interface between PyBERTHA
and PyADF. We showed that this results both in a very efficient numerical representation of
the electron density of the active system and in a straightforward
evaluation of the matrix representation in the relativistic G-spinor
basis of the embedding potential.

The accuracy and numerical stability 
of this approach, also using different auxiliary fitting basis sets, has been demonstrated 
on the simple NH$_3$-H$_2$O system. We compared the dipole moment components
and diagonal elements of the polarizability tensor of the isolated water  molecule with respect to the embedded water (i.e., NH$_3$-H$_2$O system). We performed the calculations  using both our DKS-in-DFT FDE implementation  as well as the previously implemented \textsc{Psi4-rt-PyEmbed} code.  The numerical results 
shown an evident quantitative agreement between the
two implementations. Indeed, both the variations
induced by the presence of the embedding system 
and the absolute values of both the dipole moments and polarizability show a good agreement.
Noteworthy, independently by the basis set used, the differences are
below of 0.001 a.u. and 0.01 a.u.  for the dipole moment components
and for the polarizability tensor components, respectively.  

We evaluated also the computational burden
on a series of gold clusters (Au$_n$, with $n=2,4,8$)
embedded into an increasing number of water molecules (5, 10, 20, 40
and 80 water molecules).  
We found the our implementation approximately scales
linearly both with respect to the size of the frozen surrounding environment and
the size of the active system. 
We efficiently parallelized, using OpenMP, two of the most demanding
steps on our computation, that is the computation of the numerical
representation of active system fitted density on grid,  as well as the
projection of the embedding potential onto fitting basis functions. The
results reported show that we are capable of reaching a final speedup of
31.1 and 29.8 using 32 threads for the two cited steps respectively.

Finally, we applied this new implementation to a series of Heavy (Rn)
and Super-Heavy elements (Cn, Fl, Og) embedded in a C$_{60}$ cage to
study the confinement effect induced by C$_{60}$ on their electronic structure. An analysis
of the embedding potential demonstrated that it can be well approximated
by a simple  radial potential which is marginally affected by the nature
of the central atom.  These latter results let us clearly envision a practical approach to be used to 
build model potential as a results of the FDE procedure.

\bibliography{biblionew}
\end{document}

\maketitle

\newpage

\section{Geometries and results}

\begin{table}
\begin{tabular}{cccc}
\hline
$H_2O$ &  &    & \\
Atom &  X &  Y &  Z \\ 
\hline
O  & 1.568501 &   0.105892  &  0.000005\\
H &  0.606736 &  -0.033962  & -0.000628\\
H &  1.940519 & -0.780005   & 0.000222\\
\hline
$NH_3$ &  &    & \\
Atom &  X &  Y &  Z \\ 
\hline
N &   -1.395591 &  -0.021564  &  0.000037 \\
H &   -1.629811 &   0.961096  & -0.106224\\
H  &  -1.862767 &  -0.512544  & -0.755974\\
H  &  -1.833547 &  -0.330770  & 0.862307\\
\end{tabular}
\caption{Geometry (\AA) of the adduct where water molecule is the active
system that is bound to an ammonia molecule, which instead plays the role of
the embedding environment.}
\label{tab:adductgeom}
\end{table}

\begin{table}
\begin{tabular}{rrrr}
        & \multicolumn{3}{c}{a) aug-cc-pvdz-decon}                             \\
        & \multicolumn{3}{c}{}                             \\
                 &   Free      & Emb      & $\Delta$               \\
       & \multicolumn{3}{c}{}                             \\
	\cline{2-4}  \\
$\mu_x$          &   -0.35424&  -0.49352 &  -0.13928 \\
$\mu_y$          &   -0.62075&  -0.62973 &  -0.00898 \\
$\mu_z$          &   -0.00024&  -0.00027 &  -0.00003 \\
$|\mu|$          &    0.71471&   0.80007 &   0.08536 \\
$\alpha_{xx}$    &    10.35  &  9.79     &   -0.56 \\
$\alpha_{yy}$    &    9.91   &  10.26    &   0.35\\
$\alpha_{zz}$    &    9.62   &  10.16    &   0.54\\
$\alpha_{iso}$   &    9.96   &  10.07    &   0.11\\
\end{tabular}
\caption{Dipole moment (components $\mu_x$, $\mu_y$, $\mu_z$ and
module $|\mu|$) and dipole polarizability (tensor diagonal components
$\alpha_{xx}$, $\alpha_{yy}$, $\alpha_{zz}$ and isotropic
contribution $\alpha_{iso}$) of both the isolated (free)
and embedded (emb.) water molecule. In the embedded water
molecule, an ammonia molecule is used as environment. Data
have been obtained using our new pyberthaemb  implementation
increasing the speed of light c = 10 $\cdot$ 137.036 au.
}
\label{tab:vk} 
\end{table}

\newpage

% How to generate fittin basis set
% TO BE DONE 

\section{Fitting basis set}

The exponents generation depends on the
smallest  and largest exponent in the primitive Gaussian exponents of
the chosen basis set and by a parameter (N=2,3,4) which together determine
the total number of exponents.
We recall that the auxiliary functions we employed
are grouped in s, sp, spd, spdf  and spdfg sets and that the exponents are shared within
each of these sets.
For instance, the A4$_{spdfg}$ (generated with N=4) and associated with cc-pvtz basis for the oxygen atom have 17 different
exponents with the angular part described with (5,7,5) auxiliary function
notation adopted by Calaminici et al. \cite{Calaminici2007}.  It describes five s sets together with 5
functions, seven spd sets together with 70 functions, and 5 spdfg sets
together with  functions 175.  For the hydrogen atom the A4$_{spdfg}$
fitting basis set correspond to a (2,4,3) fitting basis set. The total
number of auxiliary function for the water molecule is 544.
Using the  same  automatic generation scheme
(even if may be not optimal) we generate different sets of exponents
of reduced size depending on the teger parameter N.
Thus, for the same cc-pvdz basis,
we generate other five fitting basis sets (A2$_{s}$, A2$_{sp}$, A2$_{spd}$,
A2$_{spdfg}$ and A3$_{spdfg}$). See the definition in the following.
\begin{verbatim}
Fitting basis set: A2s

H
5
13.14560 0
16.43200 0
3.28640  0
0.82160  0
0.20540  0

O
9
7012.35200 0
8765.44000 1
1753.08800 1
438.27200  1
109.56800  1
27.39200   1
6.84800    1
1.71200    1
0.42800    1
		
Fitting basis set: A2sp

H
5
13.14560 0
16.43200 1
3.28640  1
0.82160  1
0.20540  1

O
9
7012.35200 0
8765.44000 1
1753.08800 1
438.27200  1
109.56800  1
27.39200   1
6.84800    1
1.71200    1
0.42800    1

Fitting basis set: A2spd

H
5
13.14560 0
16.43200 2
3.28640  2
0.82160  2
0.20540  2

O
7012.35200 0
8765.44000 2
1753.08800 2
438.27200  2
109.56800  2
27.39200   2
6.84800    2
1.71200    2
0.42800    2

Fitting basis set: A2spdfg

H
5
13.14560 0
16.43200 2
3.28640  2
0.82160  4
0.20540  4

O
9
7012.35200 0
8765.44000 2
1753.08800 2
438.27200  2
109.56800  2
27.39200   2
6.84800    4
1.71200    4
0.42800    4

Fitting basis set: A3spdfg

H
7
14.45364 0
21.68046 2
4.81788  2
1.60596  2
0.53532  4
0.17844  4
0.05948  4

O
11
10837.45980 0
16256.18970 2
3612.48660  2
1204.16220  2
401.38740   2
133.79580   2
44.59860    2
14.86620    4
4.95540     4
1.65180     4
0.55060     4

Fitting basis set: A4spdfg

H
8
15.61600 0
31.23200 2
7.80800  2
3.90400  2
1.95200  2
0.97600  4
0.48800  4
0.24400  4

O
17
14024.70400 0
28049.40800 2
7012.35200  2
3506.17600  2
1753.08800  2
876.54400   2
438.27200   2
219.13600   2
109.56800   2
54.78400    2
27.39200    2
13.69600    4
6.84800     4
3.42400     4
1.71200     4
0.85600     4
0.42800     4
\end{verbatim}

\newpage

\section{Computational time}
\begin{figure}[h!]
\centering
\includegraphics[scale=0.5]{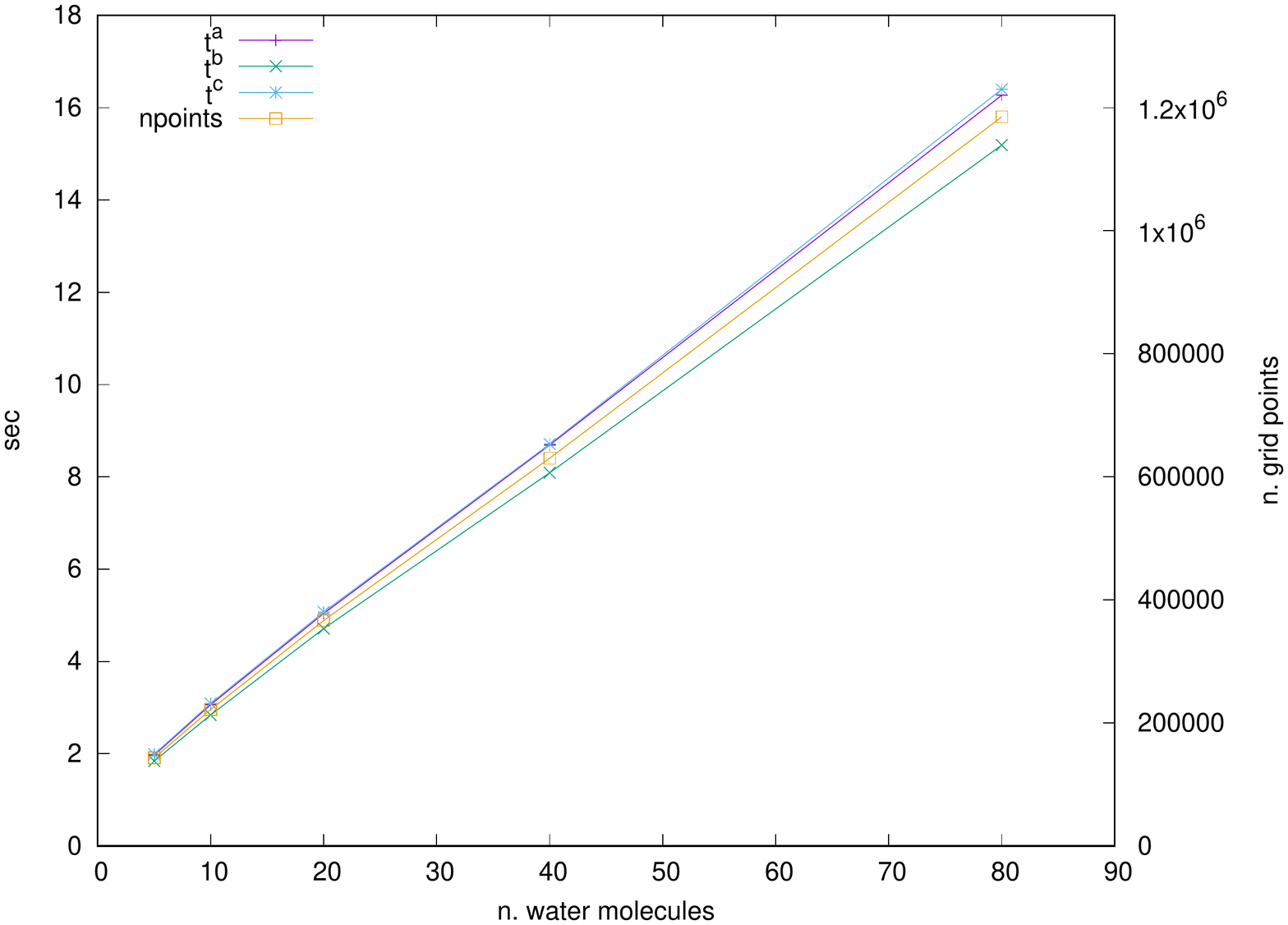}
	\caption{Computational time of the tasks associated with the FDE procedure (a,b,
	and c; see main text for the definitions) and the total number of grid points 
	are reported for Au$_4$@(H$_2$O)$_{n}$ (with n equal to 5, 10, 20, 40 and 80). 
	Time is in sec.}
\label{fig:fig1}
\end{figure}

\newpage

\section{Fitting Basis set definition for Rn, Cn, Fl and Og}
\label{fitting}

\begin{verbatim}
Rn
 25
 11.3695320E+07 0                       
 2.9358812E+07  0                       
 9.4982512E+06  0                       
 3.3357532E+06  0                       
 12.5034106E+05 2                       
 4.8936286E+05  2                       
 19.9232080E+04 2                       
 8.3919070E+04  2                       
 3.6483407E+04  2                       
 16.3264454E+03 2                       
 7.5044994E+03  2                       
 3.5359806E+03  2                       
 17.0195914E+02 2                       
 8.0823622E+02  2                       
 4.0781424E+02  2                       
 2.1070864E+02  2                       
 10.3445490E+01 2                       
 5.4675064E+01  2                       
 2.5256416E+01  2                       
 13.2706936E+00 2                       
 5.1310780E+00  4                       
 2.5895016E+00  4                       
 7.4055918E-01  4                       
 2.8874634E-01  4                       
 8.8783890E-02  4                       

Cn
 26
10.5525682E+07 0                        
2.8026742E+07  0                        
9.5036698E+06  0                        
3.5503500E+06  0                        
14.4081728E+05 2                        
6.1222764E+05  2                        
2.7032544E+05  2                        
12.2344004E+04 2                        
5.6558338E+04  2                        
2.6594564E+04  2                        
12.7103702E+03 2                        
6.1719122E+03  2                        
3.0521732E+03  2                        
15.2210270E+02 2                        
7.8252222E+02  2                        
4.1243852E+02  2                        
2.1889478E+02  2                        
11.9177684E+01 2                        
5.9408372E+01  2                        
3.3631760E+01  2                        
16.1254664E+00 2                        
8.7343090E+00  2                        
3.4773516E+00  4                        
16.6994090E-01 4                        
4.5980612E-01  4                        
16.4591954E-02 4                        

Fl
 27
10.5100528E+07 0  
2.7919038E+07  0                        
9.4708040E+06  0                        
3.5413462E+06  0                        
14.4108562E+05 2                        
6.1494150E+05  2                        
2.7307174E+05  2                        
12.4347488E+04 2                        
5.7844092E+04  2                        
2.7353736E+04  2                        
13.1373470E+03 2                        
6.4051104E+03  2                        
3.1786378E+03  2                        
15.9218248E+02 2                        
8.2096760E+02  2                        
4.3381696E+02  2                        
2.3128896E+02  2                        
12.6241812E+01 2                        
6.3432228E+01  2                        
3.6004852E+01  2                        
17.4614446E+00 2                        
9.5272666E+00  2                        
3.8762812E+00  4                        
19.0933950E-01 4                        
5.7274828E-01  4                        
2.1504274E-01  4                        
4.4258228E-02  4                        

Og
 27
10.4908686E+07 0
2.7868550E+07  0
9.4517386E+06  0
3.5331196E+06  0
14.4000862E+05 2
6.1668480E+05  2
2.7560868E+05  2
12.651001E+04  2
5.9375108E+04  2
2.8313948E+04  2
13.6997362E+03 2
6.7204318E+03  2
3.3536848E+03  2
16.9430534E+02 2
8.7817464E+02  2
4.6504244E+02  2
2.5194358E+02  2
13.8681056E+01 2
7.0841358E+01  2
4.0483622E+01  2
2.0105508E+01  2
11.1240528E+00 2
4.7031744E+00  4
2.4220012E+00  4
8.2445578E-01  4
3.1806344E-01  4
8.7080594E-02  4
\end{verbatim}

\section{Evaluation of the spherical average of the EMBP} 

As in the case of the the contour plot, reported in the main text, it is important to
underlined as the spherical average of the EMBP for the neutral endohedral fullerenes A@C$_{60}$ (A=Rn,Og,Fl,Cn) is the result of a Nearest-neighbor interpolation performed
starting from the potential represented on the original non homogeneous ADF grid. Initially we took advantage of the capability of PyADF to
dump the Embedding Potential into a file. Subsequently the file is processed via a simple Python script (i.e. \textbf{gridtodx.py} \cite{pyberthautils}) that, using a Nearest neighbor interpolation, as implemented in 
the SciPy \cite{2020SciPy-NMeth}, transforms the potential, originally represented on a non homogeneous grid to a potential evaluated on a homogeneous grid, producing a DX file via the gridDataFormats package \cite{griddata}.  Once the EMBP has been dumped in a DX file, using an homogeneously spaced grid, quite easily the Spherical overage can be evaluated (i.e. \textbf{avgalongr.py} \cite{pyberthautils}) simply considering the average of the Embedding Potential values contained in spherical shells, always centered in the A atom,  of increasing radius and constant thickness $dr$.   
\newpage

\begin{figure}
  \includegraphics[width=0.8\linewidth]{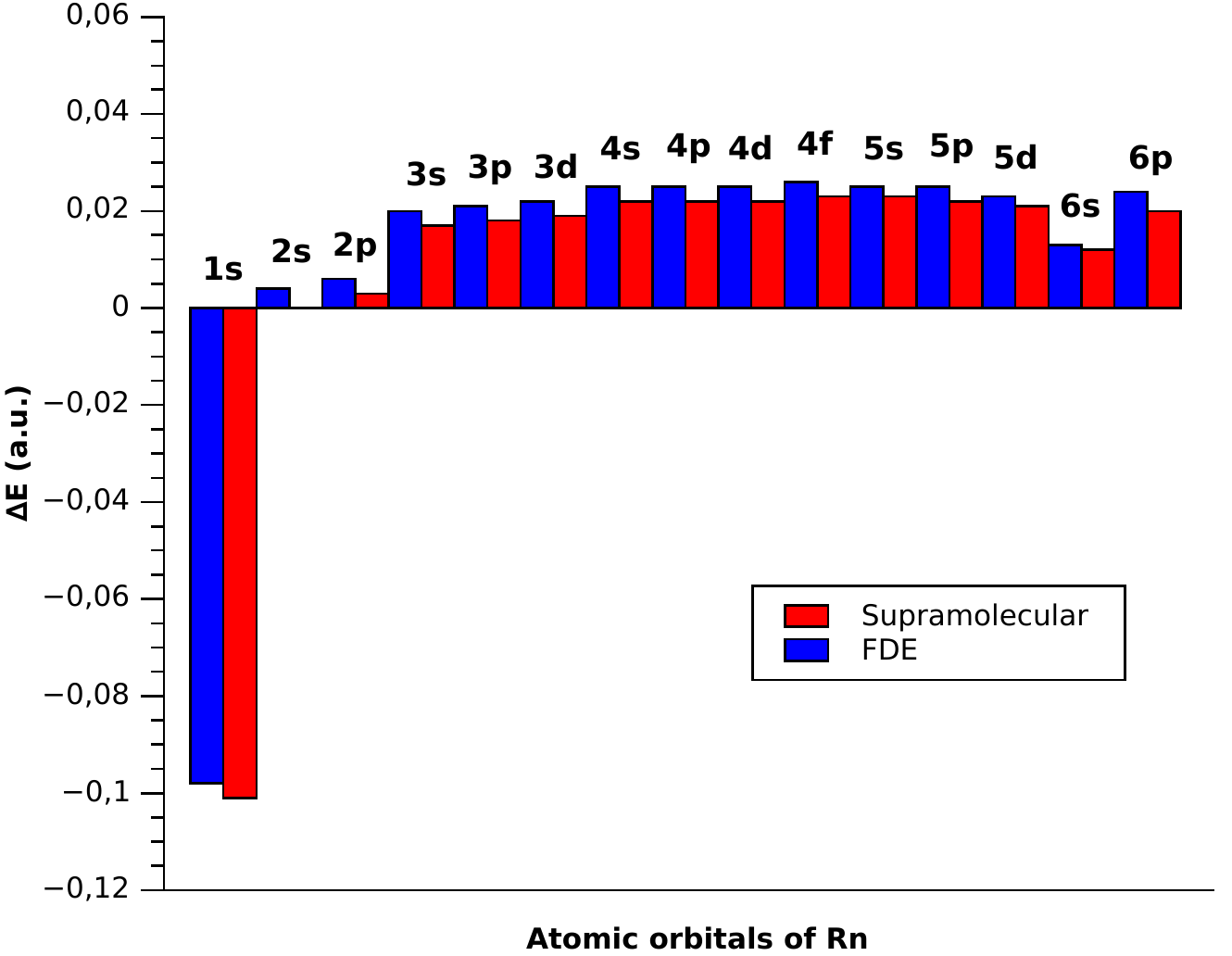}
	\caption{Differences in orbital energies with respect to the isolated Rn atom for the Rn-C$_{60}$ system using 
	the frozen density embedding (FDE) scheme (where Rn is used as active system while  C$_{60}$ as frozen environment) and the supramolecular calculation (Supramolecular). 
	All calculations have been carried out with the ADF code (2016.104 version)\cite{ADF2017authors} using the BLYP functional together with the TZP basis set, the ZORA Hamiltonian for the inclusion of scalar relativistic effects, 
	no frozen core approximation 
	and the quality of both the density fitting and the numerical integration set to "Very good". 
	The FDE calculations have been carried out using a full supramolecular grid for the frozen subsystem.}
  \label{fig:bsl}
\end{figure}

\newpage

\section{Eigenvalues for neutral endohedral fullerenes A@C$_{60}$ (A=Rn,Og,Fl,Cn)} 

\begin{longtable}{c|c|c|c|c}
\caption{Rn atom eigenvalues calculations were carried out using a basis set for the Rn atom generated by uncontracting 
triple--$\zeta$ quality Dyall's basis sets  \cite{Dyall2004,dyall2010revised,dyall2006relativistic,dyall09_12638} augmented with the related polarization and correlating functions. Final basis set schemes is (31s27p18d12f4g1h). For all the elements we used auxiliary basis sets already employed in Ref.\citenum{Rampino2015} and are explicitly reported in Section \ref{fitting}.}\\
\hline
Eigenvalue & Isolated (a.u.) &   SPM (a.u.)   & FDE C$_{60}$ (a.u.) &  EMBP Spherical   \\
           &                 &                &                     &   Average (a.u.)  \\
\hline
     1     & -3614.89008690  & -3614.91899840 & -3614.88204411      &   -3614.88244268  \\
     2     & -3614.89008690  & -3614.91899840 & -3614.88204411      &   -3614.88244268  \\
     3     &  -657.66880806  &  -657.69759019 &  -657.66072974      &    -657.66108023  \\
     4     &  -657.66880806  &  -657.69759019 &  -657.66072974      &    -657.66108023  \\
     5     &  -633.13630845  &  -633.16511925 &  -633.12823503      &    -633.12859842  \\
     6     &  -633.13630845  &  -633.16511925 &  -633.12823503      &    -633.12859842  \\
     7     &  -532.03165711  &  -532.06044938 &  -532.02357799      &    -532.02393103  \\
     8     &  -532.03165711  &  -532.06044938 &  -532.02357799      &    -532.02393103  \\
     9     &  -532.03165711  &  -532.06044925 &  -532.02357798      &    -532.02393093  \\
    10     &  -532.03165711  &  -532.06044925 &  -532.02357798      &    -532.02393093  \\
    11     &  -161.24972086  &  -161.27857149 &  -161.24161423      &    -161.24186586  \\
    12     &  -161.24972086  &  -161.27857149 &  -161.24161423      &    -161.24186586  \\
    13     &  -150.03564968  &  -150.06449386 &  -150.02754166      &    -150.02780118  \\
    14     &  -150.03564968  &  -150.06449386 &  -150.02754166      &    -150.02780118  \\
    15     &  -127.16985627  &  -127.19871977 &  -127.16175337      &    -127.16199774  \\
    16     &  -127.16985627  &  -127.19871977 &  -127.16175337      &    -127.16199774  \\
    17     &  -127.16985627  &  -127.19871943 &  -127.16175334      &    -127.16199754  \\
    18     &  -127.16985627  &  -127.19871943 &  -127.16175334      &    -127.16199754  \\
    19     &  -108.82681068  &  -108.85563632 &  -108.81870354      &    -108.81896751  \\
    20     &  -108.82681068  &  -108.85563632 &  -108.81870354      &    -108.81896751  \\
    21     &  -108.82681068  &  -108.85563594 &  -108.81870350      &    -108.81896734  \\
    22     &  -108.82681068  &  -108.85563594 &  -108.81870350      &    -108.81896734  \\
    23     &  -104.03878498  &  -104.06761653 &  -104.03067811      &    -104.03093824  \\
    24     &  -104.03878498  &  -104.06761653 &  -104.03067811      &    -104.03093824  \\
    25     &  -104.03878498  &  -104.06761619 &  -104.03067810      &    -104.03093681  \\
    26     &  -104.03878498  &  -104.06761619 &  -104.03067810      &    -104.03093681  \\
    27     &  -104.03878498  &  -104.06761482 &  -104.03067805      &    -104.03093639  \\
    28     &  -104.03878498  &  -104.06761482 &  -104.03067805      &    -104.03093639  \\
    29     &   -38.41192817  &   -38.44100309 &   -38.40395119      &     -38.40402526  \\
    30     &   -38.41192817  &   -38.44100309 &   -38.40395119      &     -38.40402526  \\
    31     &   -33.45435309  &   -33.48342696 &   -33.44637606      &     -33.44645131  \\
    32     &   -33.45435309  &   -33.48342696 &   -33.44637606      &     -33.44645131  \\
    33     &   -27.70583185  &   -27.73492964 &   -27.69787781      &     -27.69792827  \\
    34     &   -27.70583185  &   -27.73492963 &   -27.69787781      &     -27.69792827  \\
    35     &   -27.70583185  &   -27.73492935 &   -27.69787775      &     -27.69792780  \\
    36     &   -27.70583185  &   -27.73492935 &   -27.69787775      &     -27.69792780  \\
    37     &   -19.64898911  &   -19.67807877 &   -19.64103414      &     -19.64108703  \\
    38     &   -19.64898911  &   -19.67807877 &   -19.64103414      &     -19.64108703  \\
    39     &   -19.64898911  &   -19.67807796 &   -19.64103406      &     -19.64108669  \\
    40     &   -19.64898911  &   -19.67807796 &   -19.64103406      &     -19.64108669  \\
    41     &   -18.57803597  &   -18.60713766 &   -18.57008767      &     -18.57014407  \\
    42     &   -18.57803597  &   -18.60713766 &   -18.57008767      &     -18.57014407  \\
    43     &   -18.57803597  &   -18.60713371 &   -18.57008755      &     -18.57012884  \\
    44     &   -18.57803597  &   -18.60713371 &   -18.57008755      &     -18.57012884  \\
    45     &   -18.57803597  &   -18.60712103 &   -18.57008722      &     -18.57012625  \\
    46     &   -18.57803597  &   -18.60712103 &   -18.57008722      &     -18.57012625  \\
    47     &    -8.06011789  &    -8.08919929 &    -8.05215929      &      -8.05222090  \\
    48     &    -8.06011789  &    -8.08919929 &    -8.05215929      &      -8.05222090  \\
    49     &    -8.06011789  &    -8.08919622 &    -8.05215921      &      -8.05220771  \\
    50     &    -8.06011789  &    -8.08919622 &    -8.05215921      &      -8.05220771  \\
    51     &    -8.06011789  &    -8.08918498 &    -8.05215890      &      -8.05220556  \\
    52     &    -8.06011789  &    -8.08918498 &    -8.05215890      &      -8.05220556  \\
    53     &    -7.80133550  &    -7.83042125 &    -7.79338052      &      -7.79344061  \\
    54     &    -7.80133550  &    -7.83042125 &    -7.79338052      &      -7.79344061  \\
    55     &    -7.80133550  &    -7.83041899 &    -7.79338046      &      -7.79342977  \\
    56     &    -7.80133550  &    -7.83041899 &    -7.79338046      &      -7.79342977  \\
    57     &    -7.80133550  &    -7.83041078 &    -7.79338019      &      -7.79342535  \\
    58     &    -7.80133550  &    -7.83041078 &    -7.79338019      &      -7.79342535  \\
    59     &    -7.80133550  &    -7.83040393 &    -7.79338008      &      -7.79341997  \\
    60     &    -7.80133550  &    -7.83040393 &    -7.79338008      &      -7.79341997  \\
    61     &    -7.36183057  &    -7.39082933 &    -7.35407337      &      -7.35398408  \\
    62     &    -7.36183057  &    -7.39082933 &    -7.35407336      &      -7.35398408  \\
    63     &    -5.56245790  &    -5.59142512 &    -5.55473911      &      -5.55464706  \\
    64     &    -5.56245790  &    -5.59142512 &    -5.55473910      &      -5.55464706  \\
    65     &    -4.35897913  &    -4.38787434 &    -4.35134711      &      -4.35125165  \\
    66     &    -4.35897913  &    -4.38787434 &    -4.35134711      &      -4.35125165  \\
    67     &    -4.35897913  &    -4.38787319 &    -4.35134692      &      -4.35125060  \\
    68     &    -4.35897913  &    -4.38787319 &    -4.35134692      &      -4.35125060  \\
    69     &    -1.77712390  &    -1.80575462 &    -1.76981056      &      -1.76972637  \\
    70     &    -1.77712390  &    -1.80575462 &    -1.76981056      &      -1.76972637  \\
    71     &    -1.77712390  &    -1.80575251 &    -1.76981027      &      -1.76972561  \\
    72     &    -1.77712390  &    -1.80575251 &    -1.76981027      &      -1.76972561  \\
    73     &    -1.61468572  &    -1.64324492 &    -1.60745552      &      -1.60737416  \\
    74     &    -1.61468572  &    -1.64324492 &    -1.60745552      &      -1.60737416  \\
    75     &    -1.61468572  &    -1.64324255 &    -1.60745537      &      -1.60737269  \\
    76     &    -1.61468572  &    -1.64324255 &    -1.60745537      &      -1.60737269  \\
    77     &    -1.61468572  &    -1.64324207 &    -1.60745512      &      -1.60737182  \\
    78     &    -1.61468572  &    -1.64324207 &    -1.60745512      &      -1.60737182  \\
    79     &    -0.79171100  &    -0.81701374 &    -0.78769031      &      -0.78760564  \\
    80     &    -0.79171100  &    -0.81701374 &    -0.78769031      &      -0.78760564  \\
    81     &    -0.37400268  &    -0.39744765 &    -0.37107930      &      -0.37098221  \\
    82     &    -0.37400268  &    -0.39744765 &    -0.37107930      &      -0.37098221  \\
    83     &    -0.24311329  &    -0.26707774 &    -0.24019052      &      -0.24012615  \\
    84     &    -0.24311329  &    -0.26707774 &    -0.24019052      &      -0.24012615  \\
    85     &    -0.24311329  &    -0.26706790 &    -0.24018861      &      -0.24012134  \\
    86     &    -0.24311329  &    -0.26706790 &    -0.24018861      &      -0.24012134  \\
  lumo     &    -0.02215500  &    -0.14838813 &    -0.03054620      &      -0.03213182  \\
\hline
\end{longtable}

\begin{longtable}{c|c|c|c|c}
\caption{Cn atom eigenvalues calculations were carried out using a basis set for the Rn atom generated by uncontracting 
triple--$\zeta$ quality Dyall's basis sets  \cite{Dyall2004,dyall2010revised,dyall2006relativistic,dyall09_12638} augmented with the related polarization and correlating functions. Final basis set schemes is (32s29p20d14f7g2h). For all the elements we used auxiliary basis sets already employed in Ref.\citenum{Rampino2015} and are explicitly reported in Section \ref{fitting}.}\\
\hline
Eigenvalue & Isolated (a.u.) &   SPM (a.u.)    & FDE C$_{60}$ (a.u.) &  EMBP                      \\
           &                 &                 &                     &  Spherical Average (a.u.)  \\
\hline
    1      & -7032.45007220  &  -7032.44923436 & -7032.44464253      &  -7032.44500826  \\
    2      & -7032.45007220  &  -7032.44923435 & -7032.44464253      &  -7032.44500826  \\
    3      & -1428.41195749  &  -1428.41156637 & -1428.40670958      &  -1428.40705819  \\
    4      & -1428.41195749  &  -1428.41156637 & -1428.40670958      &  -1428.40705819  \\
    5      & -1393.08239459  &  -1393.08185954 & -1393.07710728      &  -1393.07746094  \\
    6      & -1393.08239459  &  -1393.08185954 & -1393.07710728      &  -1393.07746093  \\
    7      &  -994.59221688  &   -994.59189562 &  -994.58698971      &   -994.58733567  \\
    8      &  -994.59221688  &   -994.59189562 &  -994.58698971      &   -994.58733567  \\
    9      &  -994.59221688  &   -994.59189560 &  -994.58698970      &   -994.58733567  \\
   10      &  -994.59221688  &   -994.59189560 &  -994.58698970      &   -994.58733567  \\
   11      &  -382.30405390  &   -382.30482963 &  -382.29905954      &   -382.29937216  \\
   12      &  -382.30405390  &   -382.30482963 &  -382.29905954      &   -382.29937216  \\
   13      &  -364.64343793  &   -364.64407005 &  -364.63840800      &   -364.63872435  \\
   14      &  -364.64343792  &   -364.64407005 &  -364.63840800      &   -364.63872435  \\
   15      &  -268.36169388  &   -268.36274504 &  -268.35677565      &   -268.35708207  \\
   16      &  -268.36169388  &   -268.36274504 &  -268.35677565      &   -268.35708207  \\
   17      &  -268.36169388  &   -268.36274493 &  -268.35677564      &   -268.35708203  \\
   18      &  -268.36169388  &   -268.36274493 &  -268.35677564      &   -268.35708203  \\
   19      &  -240.28577774  &   -240.28650376 &  -240.28077882      &   -240.28109278  \\
   20      &  -240.28577774  &   -240.28650376 &  -240.28077882      &   -240.28109278  \\
   21      &  -240.28577774  &   -240.28650365 &  -240.28077881      &   -240.28109275  \\
   22      &  -240.28577774  &   -240.28650365 &  -240.28077881      &   -240.28109275  \\
   23      &  -223.83085096  &   -223.83169143 &  -223.82587977      &   -223.82619077  \\
   24      &  -223.83085096  &   -223.83169143 &  -223.82587977      &   -223.82619077  \\
   25      &  -223.83085096  &   -223.83169134 &  -223.82587977      &   -223.82619063  \\
   26      &  -223.83085096  &   -223.83169134 &  -223.82587977      &   -223.82619063  \\
   27      &  -223.83085096  &   -223.83169081 &  -223.82587976      &   -223.82619060  \\
   28      &  -223.83085096  &   -223.83169081 &  -223.82587976      &   -223.82619060  \\
   29      &  -108.88021773  &   -108.88292836 &  -108.87578446      &   -108.87604491  \\
   30      &  -108.88021773  &   -108.88292836 &  -108.87578446      &   -108.87604491  \\
   31      &  -100.16924690  &   -100.17188767 &  -100.16479372      &   -100.16505663  \\
   32      &  -100.16924690  &   -100.17188767 &  -100.16479372      &   -100.16505663  \\
   33      &   -72.76445385  &    -72.76761411 &   -72.76015200      &    -72.76040646  \\
   34      &   -72.76445385  &    -72.76761411 &   -72.76015200      &    -72.76040646  \\
   35      &   -72.76445385  &    -72.76761406 &   -72.76015196      &    -72.76040643  \\
   36      &   -72.76445385  &    -72.76761406 &   -72.76015196      &    -72.76040643  \\
   37      &   -59.12332404  &    -59.12636949 &   -59.11899023      &    -59.11924693  \\
   38      &   -59.12332404  &    -59.12636949 &   -59.11899023      &    -59.11924693  \\
   39      &   -59.12332404  &    -59.12636938 &   -59.11899019      &    -59.11924687  \\
   40      &   -59.12332404  &    -59.12636938 &   -59.11899019      &    -59.11924687  \\
   41      &   -54.69677467  &    -54.69995406 &   -54.69247766      &    -54.69273389  \\
   42      &   -54.69677467  &    -54.69995406 &   -54.69247766      &    -54.69273389  \\
   43      &   -54.69677467  &    -54.69995221 &   -54.69247762      &    -54.69273153  \\
   44      &   -54.69677467  &    -54.69995221 &   -54.69247762      &    -54.69273153  \\
   45      &   -54.69677467  &    -54.69994237 &   -54.69247757      &    -54.69273082  \\
   46      &   -54.69677467  &    -54.69994237 &   -54.69247757      &    -54.69273082  \\
   47      &   -36.79274450  &    -36.79567771 &   -36.78836720      &    -36.78862774  \\
   48      &   -36.79274450  &    -36.79567771 &   -36.78836720      &    -36.78862774  \\
   49      &   -36.79274450  &    -36.79567634 &   -36.78836717      &    -36.78862585  \\
   50      &   -36.79274450  &    -36.79567634 &   -36.78836717      &    -36.78862585  \\
   51      &   -36.79274450  &    -36.79566842 &   -36.78836712      &    -36.78862529  \\
   52      &   -36.79274450  &    -36.79566842 &   -36.78836712      &    -36.78862529  \\
   53      &   -35.49140797  &    -35.49440957 &   -35.48705004      &    -35.48730947  \\
   54      &   -35.49140797  &    -35.49440957 &   -35.48705004      &    -35.48730947  \\
   55      &   -35.49140797  &    -35.49440657 &   -35.48705002      &    -35.48730800  \\
   56      &   -35.49140797  &    -35.49440657 &   -35.48705002      &    -35.48730800  \\
   57      &   -35.49140797  &    -35.49440309 &   -35.48705000      &    -35.48730685  \\
   58      &   -35.49140797  &    -35.49440309 &   -35.48705000      &    -35.48730685  \\
   59      &   -35.49140797  &    -35.49439666 &   -35.48704995      &    -35.48730625  \\
   60      &   -35.49140797  &    -35.49439666 &   -35.48704995      &    -35.48730625  \\
   61      &   -28.02372229  &    -28.02842357 &   -28.01984552      &    -28.02004574  \\
   62      &   -28.02372229  &    -28.02842357 &   -28.01984552      &    -28.02004574  \\
   63      &   -24.11984009  &    -24.12457622 &   -24.11597417      &    -24.11617201  \\
   64      &   -24.11984009  &    -24.12457622 &   -24.11597417      &    -24.11617201  \\
   65      &   -16.28263942  &    -16.28787031 &   -16.27892117      &    -16.27908354  \\
   66      &   -16.28263942  &    -16.28787031 &   -16.27892117      &    -16.27908354  \\
   67      &   -16.28263942  &    -16.28786916 &   -16.27892106      &    -16.27908310  \\
   68      &   -16.28263942  &    -16.28786916 &   -16.27892106      &    -16.27908310  \\
   69      &   -10.57527237  &    -10.58063727 &   -10.57160066      &    -10.57175746  \\
   70      &   -10.57527237  &    -10.58063727 &   -10.57160066      &    -10.57175746  \\
   71      &   -10.57527237  &    -10.58063652 &   -10.57160050      &    -10.57175714  \\
   72      &   -10.57527237  &    -10.58063652 &   -10.57160050      &    -10.57175714  \\
   73      &    -9.46617915  &     -9.47168146 &    -9.46254743      &     -9.46270395  \\
   74      &    -9.46617915  &     -9.47168146 &    -9.46254743      &     -9.46270395  \\
   75      &    -9.46617915  &     -9.47167368 &    -9.46254728      &     -9.46269557  \\
   76      &    -9.46617915  &     -9.47167368 &    -9.46254728      &     -9.46269557  \\
   77      &    -9.46617915  &     -9.47163824 &    -9.46254713      &     -9.46269273  \\
   78      &    -9.46617915  &     -9.47163824 &    -9.46254713      &     -9.46269273  \\
   79      &    -4.93682232  &     -4.94335514 &    -4.93360189      &     -4.93374040  \\
   80      &    -4.93682232  &     -4.94335514 &    -4.93360189      &     -4.93374040  \\
   81      &    -3.53509878  &     -3.54172862 &    -3.53195095      &     -3.53209276  \\
   82      &    -3.53509878  &     -3.54172862 &    -3.53195095      &     -3.53209276  \\
   83      &    -2.75156393  &     -2.75740433 &    -2.74806229      &     -2.74821074  \\
   84      &    -2.75156393  &     -2.75740433 &    -2.74806229      &     -2.74821074  \\
   85      &    -2.75156393  &     -2.75739827 &    -2.74806212      &     -2.74820447  \\
   86      &    -2.75156393  &     -2.75739827 &    -2.74806212      &     -2.74820447  \\
   87      &    -2.75156393  &     -2.75737121 &    -2.74806191      &     -2.74820214  \\
   88      &    -2.75156393  &     -2.75737121 &    -2.74806191      &     -2.74820214  \\
   89      &    -2.51939601  &     -2.52529935 &    -2.51591765      &     -2.51606512  \\
   90      &    -2.51939601  &     -2.52529935 &    -2.51591765      &     -2.51606512  \\
   91      &    -2.51939601  &     -2.52528945 &    -2.51591751      &     -2.51606046  \\
   92      &    -2.51939601  &     -2.52528945 &    -2.51591751      &     -2.51606046  \\
   93      &    -2.51939601  &     -2.52527896 &    -2.51591741      &     -2.51605694  \\
   94      &    -2.51939601  &     -2.52527896 &    -2.51591741      &     -2.51605694  \\
   95      &    -2.51939601  &     -2.52525820 &    -2.51591721      &     -2.51605501  \\
   96      &    -2.51939601  &     -2.52525820 &    -2.51591721      &     -2.51605501  \\
   97      &    -1.92297586  &     -1.92981522 &    -1.92019433      &     -1.92034235  \\
   98      &    -1.92297586  &     -1.92981522 &    -1.92019433      &     -1.92034235  \\
   99      &    -1.92297586  &     -1.92981308 &    -1.92019378      &     -1.92034154  \\
  100      &    -1.92297586  &     -1.92981308 &    -1.92019378      &     -1.92034154  \\
  101      &    -0.36730789  &     -0.37391233 &    -0.36577740      &     -0.36590764  \\
  102      &    -0.36730789  &     -0.37391233 &    -0.36577740      &     -0.36590764  \\
  103      &    -0.36730789  &     -0.37391079 &    -0.36577638      &     -0.36590700  \\
  104      &    -0.36730789  &     -0.37391079 &    -0.36577638      &     -0.36590700  \\
  105      &    -0.33865992  &     -0.34563112 &    -0.33893155      &     -0.33903818  \\
  106      &    -0.33865992  &     -0.34563112 &    -0.33893155      &     -0.33903818  \\
  107      &    -0.25067057  &     -0.25725550 &    -0.24947103      &     -0.24959746  \\
  108      &    -0.25067057  &     -0.25725550 &    -0.24947103      &     -0.24959746  \\
  109      &    -0.25067057  &     -0.25725075 &    -0.24947005      &     -0.24959717  \\
  110      &    -0.25067057  &     -0.25725075 &    -0.24947005      &     -0.24959717  \\
  111      &    -0.25067057  &     -0.25723853 &    -0.24946933      &     -0.24959560  \\
  112      &    -0.25067057  &     -0.25723853 &    -0.24946933      &     -0.24959560  \\
  lumo     &    -0.09869815  &     -0.20143106 &    -0.10478383      &     -0.10490259  \\
\hline
\end{longtable}

\begin{longtable}{c|c|c|c|c}
\caption{Fl atom eigenvalues calculations were carried out using a basis set for the Rn atom generated by uncontracting 
triple--$\zeta$ quality Dyall's basis sets  \cite{Dyall2004,dyall2010revised,dyall2006relativistic,dyall09_12638} augmented with the related polarization and correlating functions. Final basis set schemes is (31s30p21d14f6g2h). For all the elements we used auxiliary basis sets already employed in Ref.\citenum{Rampino2015} and are explicitly reported in Section \ref{fitting}.}\\
\hline
Eigenvalue & Isolated (a.u.) &   SPM (a.u.)    & FDE C$_{60}$ (a.u.) &  EMBP                      \\
           &                 &                 &                     &  Spherical Average (a.u.)  \\
\hline
      1    &  -7384.82059694 &  -7384.86441378 &  -7384.81812651     &  -7384.81782805 \\
      2    &  -7384.82059694 &  -7384.86441378 &  -7384.81812651     &  -7384.81782805 \\
      3    &  -1513.68301889 &  -1513.72596392 &  -1513.68057206     &  -1513.68027823 \\
      4    &  -1513.68301889 &  -1513.72596392 &  -1513.68057206     &  -1513.68027823 \\
      5    &  -1478.65448296 &  -1478.69757441 &  -1478.65204844     &  -1478.65175426 \\
      6    &  -1478.65448296 &  -1478.69757441 &  -1478.65204844     &  -1478.65175426 \\
      7    &  -1037.05154341 &  -1037.09440831 &  -1037.04908835     &  -1037.04879453 \\
      8    &  -1037.05154341 &  -1037.09440831 &  -1037.04908835     &  -1037.04879453 \\
      9    &  -1037.05154341 &  -1037.09440813 &  -1037.04908834     &  -1037.04879451 \\
     10    &  -1037.05154341 &  -1037.09440813 &  -1037.04908834     &  -1037.04879451 \\
     11    &   -407.16520100 &   -407.20797661 &   -407.16253493     &   -407.16223615 \\
     12    &   -407.16520100 &   -407.20797661 &   -407.16253493     &   -407.16223615 \\
     13    &   -389.21052846 &   -389.25331925 &   -389.20788586     &   -389.20758722 \\
     14    &   -389.21052846 &   -389.25331925 &   -389.20788586     &   -389.20758722 \\
     15    &   -282.12499267 &   -282.16775599 &   -282.12228668     &   -282.12198839 \\
     16    &   -282.12499267 &   -282.16775599 &   -282.12228668     &   -282.12198839 \\
     17    &   -282.12499267 &   -282.16775567 &   -282.12228666     &   -282.12198835 \\
     18    &   -282.12499267 &   -282.16775567 &   -282.12228666     &   -282.12198835 \\
     19    &   -253.21978789 &   -253.26251748 &   -253.21713673     &   -253.21683868 \\
     20    &   -253.21978789 &   -253.26251748 &   -253.21713673     &   -253.21683868 \\
     21    &   -253.21978789 &   -253.26251718 &   -253.21713671     &   -253.21683864 \\
     22    &   -253.21978789 &   -253.26251718 &   -253.21713671     &   -253.21683864 \\
     23    &   -235.36191460 &   -235.40464237 &   -235.35924348     &   -235.35894541 \\
     24    &   -235.36191460 &   -235.40464237 &   -235.35924348     &   -235.35894541 \\
     25    &   -235.36191460 &   -235.40464202 &   -235.35924347     &   -235.35894507 \\
     26    &   -235.36191460 &   -235.40464202 &   -235.35924347     &   -235.35894507 \\
     27    &   -235.36191460 &   -235.40464192 &   -235.35924346     &   -235.35894504 \\
     28    &   -235.36191460 &   -235.40464192 &   -235.35924346     &   -235.35894504 \\
     29    &   -117.25255536 &   -117.29531109 &   -117.24965499     &   -117.24935632 \\
     30    &   -117.25255536 &   -117.29531109 &   -117.24965499     &   -117.24935632 \\
     31    &   -108.27676416 &   -108.31952250 &   -108.27387308     &   -108.27357481 \\
     32    &   -108.27676416 &   -108.31952250 &   -108.27387308     &   -108.27357481 \\
     33    &    -77.59480300 &    -77.63755002 &    -77.59184571     &    -77.59155651 \\
     34    &    -77.59480300 &    -77.63755002 &    -77.59184571     &    -77.59155651 \\
     35    &    -77.59480300 &    -77.63754946 &    -77.59184568     &    -77.59155651 \\
     36    &    -77.59480300 &    -77.63754946 &    -77.59184568     &    -77.59155651 \\
     37    &    -63.48169916 &    -63.52443735 &    -63.47875824     &    -63.47846788 \\
     38    &    -63.48169916 &    -63.52443735 &    -63.47875824     &    -63.47846788 \\
     39    &    -63.48169916 &    -63.52443680 &    -63.47875822     &    -63.47846779 \\
     40    &    -63.48169916 &    -63.52443680 &    -63.47875822     &    -63.47846779 \\
     41    &    -58.62892086 &    -58.67165757 &    -58.62596274     &    -58.62567909 \\
     42    &    -58.62892086 &    -58.67165757 &    -58.62596274     &    -58.62567909 \\
     43    &    -58.62892086 &    -58.67165745 &    -58.62596268     &    -58.62567291 \\
     44    &    -58.62892086 &    -58.67165745 &    -58.62596268     &    -58.62567291 \\
     45    &    -58.62892086 &    -58.67165344 &    -58.62596264     &    -58.62567227 \\
     46    &    -58.62892086 &    -58.67165344 &    -58.62596264     &    -58.62567227 \\
     47    &    -40.13178199 &    -40.17451891 &    -40.12884579     &    -40.12855744 \\
     48    &    -40.13178199 &    -40.17451891 &    -40.12884579     &    -40.12855744 \\
     49    &    -40.13178199 &    -40.17451882 &    -40.12884574     &    -40.12855252 \\
     50    &    -40.13178199 &    -40.17451882 &    -40.12884574     &    -40.12855252 \\
     51    &    -40.13178199 &    -40.17451563 &    -40.12884570     &    -40.12855207 \\
     52    &    -40.13178199 &    -40.17451563 &    -40.12884570     &    -40.12855207 \\
     53    &    -38.69576358 &    -38.73850113 &    -38.69281910     &    -38.69253245 \\
     54    &    -38.69576358 &    -38.73850113 &    -38.69281910     &    -38.69253245 \\
     55    &    -38.69576358 &    -38.73850046 &    -38.69281904     &    -38.69252824 \\
     56    &    -38.69576358 &    -38.73850046 &    -38.69281904     &    -38.69252824 \\
     57    &    -38.69576358 &    -38.73849845 &    -38.69281902     &    -38.69252681 \\
     58    &    -38.69576358 &    -38.73849845 &    -38.69281902     &    -38.69252681 \\
     59    &    -38.69576358 &    -38.73849658 &    -38.69281899     &    -38.69252463 \\
     60    &    -38.69576358 &    -38.73849658 &    -38.69281899     &    -38.69252463 \\
     61    &    -30.89239614 &    -30.93511873 &    -30.88923014     &    -30.88895927 \\
     62    &    -30.89239614 &    -30.93511873 &    -30.88923014     &    -30.88895927 \\
     63    &    -26.81395743 &    -26.85667911 &    -26.81078844     &    -26.81051672 \\
     64    &    -26.81395743 &    -26.85667911 &    -26.81078844     &    -26.81051672 \\
     65    &    -17.91416501 &    -17.95686734 &    -17.91093432     &    -17.91064622 \\
     66    &    -17.91416501 &    -17.95686734 &    -17.91093432     &    -17.91064622 \\
     67    &    -17.91416501 &    -17.95686620 &    -17.91093427     &    -17.91064588 \\
     68    &    -17.91416501 &    -17.95686620 &    -17.91093427     &    -17.91064588 \\
     69    &    -11.93662921 &    -11.97930797 &    -11.93338985     &    -11.93309840 \\
     70    &    -11.93662921 &    -11.97930797 &    -11.93338985     &    -11.93309840 \\
     71    &    -11.93662921 &    -11.97930672 &    -11.93338983     &    -11.93309831 \\
     72    &    -11.93662921 &    -11.97930672 &    -11.93338983     &    -11.93309831 \\
     73    &    -10.69339104 &    -10.73606051 &    -10.69014066     &    -10.68985912 \\
     74    &    -10.69339104 &    -10.73606051 &    -10.69014066     &    -10.68985912 \\
     75    &    -10.69339104 &    -10.73605948 &    -10.69014041     &    -10.68983896 \\
     76    &    -10.69339104 &    -10.73605948 &    -10.69014041     &    -10.68983896 \\
     77    &    -10.69339104 &    -10.73604667 &    -10.69014034     &    -10.68983653 \\
     78    &    -10.69339104 &    -10.73604667 &    -10.69014034     &    -10.68983653 \\
     79    &     -5.80073609 &     -5.84328165 &     -5.79743355     &     -5.79713126 \\
     80    &     -5.80073609 &     -5.84328165 &     -5.79743355     &     -5.79713126 \\
     81    &     -4.28884501 &     -4.33140900 &     -4.28555277     &     -4.28525578 \\
     82    &     -4.28884501 &     -4.33140900 &     -4.28555277     &     -4.28525578 \\
     83    &     -3.60491676 &     -3.64754878 &     -3.60164741     &     -3.60136013 \\
     84    &     -3.60491676 &     -3.64754878 &     -3.60164741     &     -3.60136013 \\
     85    &     -3.60491676 &     -3.64754854 &     -3.60164720     &     -3.60134440 \\
     86    &     -3.60491676 &     -3.64754854 &     -3.60164720     &     -3.60134440 \\
     87    &     -3.60491676 &     -3.64753820 &     -3.60164718     &     -3.60134265 \\
     88    &     -3.60491676 &     -3.64753820 &     -3.60164718     &     -3.60134265 \\
     89    &     -3.33398980 &     -3.37661888 &     -3.33071714     &     -3.33043068 \\
     90    &     -3.33398980 &     -3.37661888 &     -3.33071714     &     -3.33043068 \\
     91    &     -3.33398980 &     -3.37661688 &     -3.33071701     &     -3.33041833 \\
     92    &     -3.33398980 &     -3.37661688 &     -3.33071701     &     -3.33041833 \\
     93    &     -3.33398980 &     -3.37661100 &     -3.33071689     &     -3.33041430 \\
     94    &     -3.33398980 &     -3.37661100 &     -3.33071689     &     -3.33041430 \\
     95    &     -3.33398980 &     -3.37660554 &     -3.33071686     &     -3.33040787 \\
     96    &     -3.33398980 &     -3.37660554 &     -3.33071686     &     -3.33040787 \\
     97    &     -2.35606244 &     -2.39862058 &     -2.35292338     &     -2.35264626 \\
     98    &     -2.35606244 &     -2.39862058 &     -2.35292338     &     -2.35264626 \\
     99    &     -2.35606244 &     -2.39861653 &     -2.35292331     &     -2.35264616 \\
    100    &     -2.35606244 &     -2.39861653 &     -2.35292331     &     -2.35264616 \\
    101    &     -0.61967527 &     -0.66114973 &     -0.61738671     &     -0.61711598 \\
    102    &     -0.61967527 &     -0.66114973 &     -0.61738671     &     -0.61711598 \\
    103    &     -0.61967527 &     -0.66114357 &     -0.61738646     &     -0.61711547 \\
    104    &     -0.61967527 &     -0.66114357 &     -0.61738646     &     -0.61711547 \\
    105    &     -0.47926140 &     -0.51603299 &     -0.47925086     &     -0.47898960 \\
    106    &     -0.47926140 &     -0.51603299 &     -0.47925086     &     -0.47898960 \\
    107    &     -0.46293263 &     -0.50379743 &     -0.46099381     &     -0.46073029 \\
    108    &     -0.46293263 &     -0.50379743 &     -0.46099381     &     -0.46073029 \\
    109    &     -0.46293263 &     -0.50378996 &     -0.46099352     &     -0.46072341 \\
    110    &     -0.46293263 &     -0.50378996 &     -0.46099352     &     -0.46072341 \\
    111    &     -0.46293263 &     -0.50378741 &     -0.46099333     &     -0.46072233 \\
    112    &     -0.46293263 &     -0.50378741 &     -0.46099333     &     -0.46072233 \\
    113    &     -0.17795820 &     -0.21854531 &     -0.17662058     &     -0.17646775 \\
    114    &     -0.17795820 &     -0.21854531 &     -0.17662058     &     -0.17646774 \\
  lumo     &     -0.05245587 &     -0.14629461 &     -0.06614950     &     -0.06600113 \\
\hline
\end{longtable}

\begin{longtable}{c|c|c|c|c}
\caption{Og atom eigenvalues calculations were carried out using a basis set for the Rn atom generated by uncontracting 
triple--$\zeta$ quality Dyall's basis sets  \cite{Dyall2004,dyall2010revised,dyall2006relativistic,dyall09_12638} augmented with the related polarization and correlating functions. Final basis set schemes is (31s30p21d14f6g2h). For all the elements we used auxiliary basis sets already employed in Ref.\citenum{Rampino2015} and are explicitly reported in Section \ref{fitting}.}\\
\hline
Eigenvalue & Isolated (a.u.) &   SPM (a.u.)    & FDE C$_{60}$ (a.u.) &  EMBP                      \\
           &                 &                 &                     &  Spherical Average (a.u.)  \\
\hline
      1    & -8143.60941658  & -8143.66233440 & -8143.59921010  & -8143.59871400 \\
      2    & -8143.60941658  & -8143.66233440 & -8143.59921010  & -8143.59871400 \\
      3    & -1700.86804673  & -1700.92078516 & -1700.85776874  & -1700.85727112 \\
      4    & -1700.86804673  & -1700.92078516 & -1700.85776874  & -1700.85727112 \\
      5    & -1668.29159261  & -1668.34442448 & -1668.28132288  & -1668.28082556 \\
      6    & -1668.29159261  & -1668.34442448 & -1668.28132288  & -1668.28082556 \\
      7    & -1125.16260496  & -1125.21532797 & -1125.15230582  & -1125.15180752 \\
      8    & -1125.16260496  & -1125.21532797 & -1125.15230582  & -1125.15180752 \\
      9    & -1125.16260496  & -1125.21532783 & -1125.15230580  & -1125.15180743 \\
     10    & -1125.16260496  & -1125.21532783 & -1125.15230580  & -1125.15180743 \\
     11    &  -461.86386763  &  -461.91643716 &  -461.85342109  &  -461.85291813 \\
     12    &  -461.86386763  &  -461.91643716 &  -461.85342109  &  -461.85291813 \\
     13    &  -443.71587653  &  -443.76846448 &  -443.70544946  &  -443.70494673 \\
     14    &  -443.71587653  &  -443.76846448 &  -443.70544946  &  -443.70494673 \\
     15    &  -311.15028068  &  -311.20283821 &  -311.13979392  &  -311.13929226 \\
     16    &  -311.15028068  &  -311.20283821 &  -311.13979392  &  -311.13929226 \\
     17    &  -311.15028068  &  -311.20283793 &  -311.13979389  &  -311.13929207 \\
     18    &  -311.15028068  &  -311.20283793 &  -311.13979389  &  -311.13929207 \\
     19    &  -280.55110644  &  -280.60366650 &  -280.54066275  &  -280.54016058 \\
     20    &  -280.55110644  &  -280.60366650 &  -280.54066275  &  -280.54016058 \\
     21    &  -280.55110644  &  -280.60366624 &  -280.54066273  &  -280.54016041 \\
     22    &  -280.55110644  &  -280.60366624 &  -280.54066273  &  -280.54016041 \\
     23    &  -259.61197021  &  -259.66452095 &  -259.60151002  &  -259.60100805 \\
     24    &  -259.61197021  &  -259.66452095 &  -259.60151002  &  -259.60100805 \\
     25    &  -259.61197021  &  -259.66452014 &  -259.60151001  &  -259.60100762 \\
     26    &  -259.61197021  &  -259.66452014 &  -259.60151001  &  -259.60100762 \\
     27    &  -259.61197021  &  -259.66451984 &  -259.60150998  &  -259.60100749 \\
     28    &  -259.61197021  &  -259.66451984 &  -259.60150998  &  -259.60100749 \\
     29    &  -135.92353208  &  -135.97601471 &  -135.91289556  &  -135.91239629 \\
     30    &  -135.92353208  &  -135.97601471 &  -135.91289556  &  -135.91239629 \\
     31    &  -126.51483911  &  -126.56732858 &  -126.50421076  &  -126.50371197 \\
     32    &  -126.51483911  &  -126.56732858 &  -126.50421076  &  -126.50371197 \\
     33    &   -88.06705420  &   -88.11950988 &   -88.05636705  &   -88.05587612 \\
     34    &   -88.06705420  &   -88.11950988 &   -88.05636705  &   -88.05587612 \\
     35    &   -88.06705420  &   -88.11950947 &   -88.05636698  &   -88.05587588 \\
     36    &   -88.06705420  &   -88.11950947 &   -88.05636698  &   -88.05587588 \\
     37    &   -72.99161955  &   -73.04407872 &   -72.98094594  &   -72.98045433 \\
     38    &   -72.99161955  &   -73.04407872 &   -72.98094594  &   -72.98045433 \\
     39    &   -72.99161955  &   -73.04407823 &   -72.98094589  &   -72.98045401 \\
     40    &   -72.99161955  &   -73.04407823 &   -72.98094589  &   -72.98045401 \\
     41    &   -67.19151145  &   -67.24397130 &   -67.18082370  &   -67.18033966 \\
     42    &   -67.19151145  &   -67.24397130 &   -67.18082370  &   -67.18033966 \\
     43    &   -67.19151145  &   -67.24395544 &   -67.18082361  &   -67.18033220 \\
     44    &   -67.19151145  &   -67.24395544 &   -67.18082361  &   -67.18033220 \\
     45    &   -67.19151145  &   -67.24395482 &   -67.18082354  &   -67.18033104 \\
     46    &   -67.19151145  &   -67.24395482 &   -67.18082354  &   -67.18033104 \\
     47    &   -47.49861849  &   -47.55108521 &   -47.48795031  &   -47.48746172 \\
     48    &   -47.49861849  &   -47.55108521 &   -47.48795031  &   -47.48746172 \\
     49    &   -47.49861849  &   -47.55107297 &   -47.48795024  &   -47.48745587 \\
     50    &   -47.49861849  &   -47.55107297 &   -47.48795024  &   -47.48745587 \\
     51    &   -47.49861849  &   -47.55107234 &   -47.48795018  &   -47.48745507 \\
     52    &   -47.49861849  &   -47.55107234 &   -47.48795018  &   -47.48745507 \\
     53    &   -45.76283759  &   -45.81530302 &   -45.75216247  &   -45.75167576 \\
     54    &   -45.76283759  &   -45.81530302 &   -45.75216247  &   -45.75167576 \\
     55    &   -45.76283759  &   -45.81529122 &   -45.75216243  &   -45.75167138 \\
     56    &   -45.76283759  &   -45.81529122 &   -45.75216243  &   -45.75167138 \\
     57    &   -45.76283759  &   -45.81529061 &   -45.75216237  &   -45.75166837 \\
     58    &   -45.76283759  &   -45.81529061 &   -45.75216237  &   -45.75166837 \\
     59    &   -45.76283759  &   -45.81528362 &   -45.75216231  &   -45.75166655 \\
     60    &   -45.76283759  &   -45.81528362 &   -45.75216231  &   -45.75166655 \\
     61    &   -37.53884950  &   -37.59116109 &   -37.52802571  &   -37.52756092 \\
     62    &   -37.53884950  &   -37.59116109 &   -37.52802571  &   -37.52756092 \\
     63    &   -33.13104298  &   -33.18335364 &   -33.12021840  &   -33.11975295 \\
     64    &   -33.13104298  &   -33.18335364 &   -33.12021840  &   -33.11975295 \\
     65    &   -21.67975586  &   -21.73199441 &   -21.66889172  &   -21.66841096 \\
     66    &   -21.67975586  &   -21.73199441 &   -21.66889172  &   -21.66841096 \\
     67    &   -21.67975586  &   -21.73199387 &   -21.66889155  &   -21.66841082 \\
     68    &   -21.67975586  &   -21.73199387 &   -21.66889155  &   -21.66841082 \\
     69    &   -15.14881823  &   -15.20103315 &   -15.13795297  &   -15.13746947 \\
     70    &   -15.14881823  &   -15.20103315 &   -15.13795297  &   -15.13746947 \\
     71    &   -15.14881823  &   -15.20103248 &   -15.13795284  &   -15.13746910 \\
     72    &   -15.14881823  &   -15.20103248 &   -15.13795284  &   -15.13746910 \\
     73    &   -13.60019245  &   -13.65241183 &   -13.58932298  &   -13.58884904 \\
     74    &   -13.60019245  &   -13.65241183 &   -13.58932298  &   -13.58884904 \\
     75    &   -13.60019245  &   -13.65236864 &   -13.58932272  &   -13.58882915 \\
     76    &   -13.60019245  &   -13.65236864 &   -13.58932272  &   -13.58882915 \\
     77    &   -13.60019245  &   -13.65236770 &   -13.58932255  &   -13.58882553 \\
     78    &   -13.60019245  &   -13.65236770 &   -13.58932255  &   -13.58882553 \\
     79    &    -8.01197874  &    -8.06381481 &    -8.00117137  &    -8.00066332 \\
     80    &    -8.01197874  &    -8.06381481 &    -8.00117137  &    -8.00066332 \\
     81    &    -6.27745122  &    -6.32923085 &    -6.26667305  &    -6.26616767 \\
     82    &    -6.27745122  &    -6.32923085 &    -6.26667305  &    -6.26616767 \\
     83    &    -5.76088057  &    -5.81301519 &    -5.75002052  &    -5.74953948 \\
     84    &    -5.76088057  &    -5.81301519 &    -5.75002052  &    -5.74953948 \\
     85    &    -5.76088057  &    -5.81298110 &    -5.75002032  &    -5.74952348 \\
     86    &    -5.76088057  &    -5.81298110 &    -5.75002032  &    -5.74952348 \\
     87    &    -5.76088057  &    -5.81297997 &    -5.75002017  &    -5.74952090 \\
     88    &    -5.76088057  &    -5.81297997 &    -5.75002017  &    -5.74952090 \\
     89    &    -5.40095224  &    -5.45307370 &    -5.39009381  &    -5.38961280 \\
     90    &    -5.40095224  &    -5.45307370 &    -5.39009381  &    -5.38961280 \\
     91    &    -5.40095224  &    -5.45304449 &    -5.39009373  &    -5.38960178 \\
     92    &    -5.40095224  &    -5.45304449 &    -5.39009373  &    -5.38960178 \\
     93    &    -5.40095224  &    -5.45304297 &    -5.39009357  &    -5.38959443 \\
     94    &    -5.40095224  &    -5.45304297 &    -5.39009357  &    -5.38959443 \\
     95    &    -5.40095224  &    -5.45302571 &    -5.39009342  &    -5.38958984 \\
     96    &    -5.40095224  &    -5.45302571 &    -5.39009342  &    -5.38958984 \\
     97    &    -3.54080603  &    -3.59227005 &    -3.53026532  &    -3.52978029 \\
     98    &    -3.54080603  &    -3.59227005 &    -3.53026532  &    -3.52978029 \\
     99    &    -3.54080603  &    -3.59226883 &    -3.53026505  &    -3.52977957 \\
    100    &    -3.54080603  &    -3.59226883 &    -3.53026505  &    -3.52977957 \\
    101    &    -1.43134111  &    -1.48222524 &    -1.42136921  &    -1.42089752 \\
    102    &    -1.43134111  &    -1.48222524 &    -1.42136921  &    -1.42089752 \\
    103    &    -1.43134111  &    -1.48222363 &    -1.42136891  &    -1.42089641 \\
    104    &    -1.43134111  &    -1.48222363 &    -1.42136891  &    -1.42089641 \\
    105    &    -1.17572159  &    -1.22628397 &    -1.16604657  &    -1.16558328 \\
    106    &    -1.17572159  &    -1.22628397 &    -1.16604657  &    -1.16558328 \\
    107    &    -1.17572159  &    -1.22627266 &    -1.16604647  &    -1.16557643 \\
    108    &    -1.17572159  &    -1.22627266 &    -1.16604647  &    -1.16557643 \\
    109    &    -1.17572159  &    -1.22627081 &    -1.16604616  &    -1.16557602 \\
    110    &    -1.17572159  &    -1.22627081 &    -1.16604616  &    -1.16557602 \\
    111    &    -0.98522833  &    -1.03292673 &    -0.97786391  &    -0.97740888 \\
    112    &    -0.98522833  &    -1.03292673 &    -0.97786391  &    -0.97740888 \\
    113    &    -0.53127218  &    -0.57611273 &    -0.52526963  &    -0.52482878 \\
    114    &    -0.53127218  &    -0.57611273 &    -0.52526963  &    -0.52482878 \\
    115    &    -0.19681621  &    -0.23992686 &    -0.19028724  &    -0.18996845 \\
    116    &    -0.19681621  &    -0.23992686 &    -0.19028724  &    -0.18996845 \\
    117    &    -0.19681621  &    -0.23991441 &    -0.19028440  &    -0.18996091 \\ 
    118    &    -0.19681621  &    -0.23991441 &    -0.19028440  &    -0.18996091 \\
   lumo    &    -0.04877619  &    -0.16891976 &    -0.05044183  &    -0.05066086 \\

\hline
\end{longtable}

\bibliography{biblionew}